\documentclass[12pt]{article} % use larger type; default would be 10pt

\UseRawInputEncoding

\usepackage[utf8]{inputenc} % set input encoding (not needed with XeLaTeX)
\usepackage{amsmath}
\usepackage{amssymb}
\usepackage{amsthm}
\usepackage[noblocks,affil-it]{authblk}
\usepackage{bm}
\usepackage{enumitem}
\usepackage{url}
\usepackage[toc,page]{appendix}
\usepackage[round]{natbib}
\usepackage{multirow}
\usepackage{color, colortbl}
\usepackage{placeins}
\usepackage{pifont}
\usepackage{makecell}
\usepackage{hyperref}
\usepackage{geometry} % to change the page dimensions
\usepackage{graphicx} % support the \includegraphics command and options

\usepackage{booktabs} % for much better looking tables
\usepackage{array} % for better arrays (eg matrices) in maths
\usepackage{verbatim} % adds environment for commenting out blocks of text & for better verbatim
\usepackage{subfig} % make it possible to include more than one captioned figure/table in a single float
\usepackage{fancyhdr} % This should be set AFTER setting up the page geometry
\usepackage[nottoc,notlof,notlot]{tocbibind} % Put the bibliography in the ToC
\usepackage[titles,subfigure]{tocloft} % Alter the style of the Table of Contents

\newcommand{\ys}{\mathbf y}
\newcommand{\yy}{\mathbf Y}
\newcommand{\Xs}{\mathbf X}
\newcommand{\xj}{\mathbf x^{(j)} }
\newcommand{\Ws}{\mathbf W}
\newcommand{\ston}{s_1 , ..., s_n}

\newcommand{\sptonp}{s_1' , ..., s_{n'}'}
\newcommand{\betas}{\bm\beta}
\newcommand{\epss}{\bm \varepsilon}

\newcommand{\argmin}[1]{\underset{#1}{\operatorname{arg}\,\operatorname{min}}\;}

\newtheorem*{remark}{Remark}

\theoremstyle{definition}

\definecolor{Gray}{gray}{0.7}
\definecolor{lightGray}{gray}{0.8}

\newbox{\myorcidaffilbox}
\sbox{\myorcidaffilbox}{\large\includegraphics[height=1.7ex]{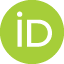}}
\newcommand{\orcid}[1]{\href{https://orcid.org/#1}{\usebox{\myorcidaffilbox}}}

\providecommand{\keywords}[1]{\textit{Keywords:} #1}

\newcommand{\revone}[1]{\textcolor{black}{#1}}

\title{\protect Maximum Likelihood Estimation of Spatially Varying Coefficient Models for Large Data with an Application to Real Estate Price Prediction}

\author[1,2]{Jakob A.\ Dambon\thanks{Email: \href{mailto:jakob.dambon@math.uzh.ch}{jakob.dambon@math.uzh.ch}, Address: Department of Mathematics, University of Zurich, Winterthurerstrasse 190, 8057 Zurich, Switzerland.}\orcid{0000-0001-5855-2017}}
\author[2]{Fabio Sigrist\thanks{Email: \href{mailto:fabio.sigrist@hslu.ch}{fabio.sigrist@hslu.ch}, Address: Institute of Financial Services Zug, Campus Zug-Rotkreuz, Suurstoffi 1, 6343 Rotkreuz, Switzerland.}\orcid{0000-0002-3994-2244}}
\author[1,3]{Reinhard Furrer\thanks{Email: \href{mailto:reinhard.furrer@math.uzh.ch}{reinhard.furrer@math.uzh.ch}, Address: Department of Mathematics, University of Zurich, Winterthurerstrasse 190, 8057 Zurich, Switzerland. \\ \\ \revone{\emph{Paper submitted to Spatial Statistics on April 3, 2020. Revised on July 30, 2020. Accepted on September 5, 2020.} \url{https://doi.org/10.1016/j.spasta.2020.100470}}}\orcid{0000-0002-6319-2332}}
\affil[1]{Department of Mathematics, University of Zurich}
\affil[2]{Institute of Financial Services Zug, Lucerne University of Applied Sciences and Arts}
\affil[3]{Department of Computational Science, University of Zurich}

\date{November 12, 2020} % Activate to display a given date or no date (if empty),
\begin{document}
\maketitle

\begin{abstract}
   In regression models for spatial data, it is often assumed that the marginal effects of covariates on the response are constant over space. In practice, this assumption might often be questionable. In this article, we show how a Gaussian process-based spatially varying coefficient (SVC) model can be estimated using maximum likelihood estimation (MLE). In addition, we present an approach that scales to large data by applying covariance tapering. We compare our methodology to existing methods such as a Bayesian approach using the stochastic partial differential equation (SPDE) link, geographically weighted regression (GWR), and eigenvector spatial filtering (ESF) in both a simulation study and an application where the goal is to predict prices of real estate apartments in Switzerland. The results from both the simulation study and application show that the MLE approach results in increased predictive accuracy and more precise estimates. Since we use a model-based approach, we can also provide predictive variances. In contrast to existing model-based approaches, our method scales better to data where both the number of spatial points is large and the number of spatially varying covariates is moderately-sized, e.g., above ten.
\end{abstract}


\section{Introduction} % literature and methods overview

Over the last years, affordable measuring techniques lead to an abundance of spatial data; not only areal but in particular spatial points data. Often, the data sets contain covariates in addition to the main variable(s) of interest. These are then used in a regression model, where the goal is either predicting the response variable or inferring the relationship between the response variable and the covariates. With further advances to reduce computational cost, we are now able to analyze large spatial data sets and the literature on models and methods on how to do so is extensive. See \citet{Cressie2011} and \citet{Heaton2019} for an overview. However, the vast majority of models assumes that covariate effects are constant over space which is not necessarily plausible.

\sloppy\emph{Spatially varying coefficient (SVC)} models allow for marginal effects to be non-stationary over space and thus offer a higher degree of flexibility. At the same time, SVC models have the advantage that they are easily interpretable. Several methodologies and applications with SVC models have been published. To name two, \emph{geographically-weighted regression (GWR)} by \citet{GWR2002} and a Bayesian framework with \emph{SVC processes} by \citet{SVC2003} are prominent examples. An application that uses both methodologies can be found in \citet{Wheeler2009} who model crime records in Houston, Texas. In a simulation study \citet{Wheeler2007} conclude that SVC processes provide more accurate regression coefficient estimates than GWR. A further comparison of GWR and SVC processes is given by \citet{Finley2010} on ecological data. It is shown that SVC processes generally have better predictive performance.

However, when it comes to estimating SVC models on large data, most of the established methodologies run into problems. Currently available implementations of Bayesian approaches such as \emph{Gaussian predictive processes} presented in \citet{Banerjee2008} or \emph{Gaussian Markov random field} approximations by \citet{Lindgren2011} are either limited by the number of SVCs within a model or the number of observations. This also holds true for the before mentioned SVC processes by \citet{SVC2003}. Finally, GWR lacks a statistical sound definition and should be regarded as a purely exploratory tool. Therefore, a geostatistical estimation and prediction method is needed that, on the one hand, can deal with large number of observations and, on the other hand, can be applied to models including many SVCs.

The outline of this article is as follows: In the next section, we introduce the data set and give a first exploratory analysis that motivates the usage of SVC models. In Section~\ref{sec:SVCmodels} we formally define SVC models and give an overview of existing methods. We motivate our approach by listing potential shortcomings of the existing methods. In Section~\ref{sec:MLE} we describe our \emph{maximum likelihood estimation (MLE)} method in detail. Section~\ref{sec:sim_study} compares the existing methods to MLE in a large simulation study on synthetic data. A further comparison on the real data set is given in Section~\ref{sec:CHap}. We summarize our findings in Section~\ref{sec:conclusion}.

\section{Data Set}\label{sec:data} % show hypotheses with existung methods like ESF / GWR and shortcomings of these 

The data set provided by Fahrl\"ander Partner (Zurich, Switzerland) consists of apartment transactions in Switzerland containing the selling price, six covariates and approximate coordinates for each transaction. The goal is to regress the selling price on the given covariates. An overview and description of the data are given in Table~\ref{tab:covariates}. 

The Swiss banking secrecy prevents disclosing the exact locations of the apartments. \revone{That is why all observations are first grouped according to a dense grid consisting of $5{,}379$ relatively small cells over Switzerland with a higher resolution in densely populated areas. The cell sizes range from $3.68$ to $59{,}988.20$ acres ($0.015$ to $242.764\ \textnormal{km}^2$), the median being $872.49$ acres ($3.531\ \textnormal{km}^2$). Instead of the exact location, we then only observe the centroid of the corresponding cell for every apartment. The easting and northing of these centroids are given in the \emph{LV03} coordinate reference system and their corresponding units are meters \citep{LV03}.}
\begin{table}
	\centering
	\caption{Overview of the response, covariates and coordinates in our data set. There are $24{,}816$ transaction records in total.} \label{tab:covariates}	
	\begin{tabular}{ p{0.18\textwidth}  p{0.57\textwidth}  p{0.15\textwidth}}
		
		\hline
		Name & Description & Range \\
		\hline
		Price    & Transaction amount in Swiss Francs (CHF) & \makecell[l]{$60 \times 10^3$ --\\ $7{,}500 \times 10^3$}\\
		\rowcolor{lightGray}
		Area       & Area in square meters & 20 -- 310\\
		\makecell[l]{Year of \\ construction} & Apartments built before 1920 are set to 1920. & 1920 -- 2017 \\
		\rowcolor{lightGray}
		\makecell[l]{Micro location \\ rating} & \makecell[l]{Rating of the location on small scale,\\ i.e., walking distance (higher meaning better)}& 1 -- 5 \\
		\makecell[l]{Standard \\ rating} & \makecell[l]{Rating of standard of the apartment \\ (higher meaning better)}& 1 -- 5 \\
		\rowcolor{lightGray}
		\makecell[l]{Renovation \\ rating} & \makecell[l]{Need for renovation \\ (lower meaning better)} & 0 -- 4\\
		Date       & Quarter in which the transaction took place & \makecell[l]{Q3 2015 -- \\ Q4 2017}\\
		\rowcolor{lightGray} 
		Easting &  & {\makecell[l]{$200\times 10^3$ --\\ $800\times 10^3$}} \\
		Northing &  & {\makecell[l]{$100\times 10^3$ --\\ $400\times 10^3$}}\\
		 \hline
	\end{tabular}
\end{table}

\subsection{Motivation and Exploratory Analysis}\label{sec:motivation}

In real estate mass appraisal, there are several works that investigate or model non-stationary covariates effects. \citet{SVC2003} used a Bayesian SVC model with coefficients defined as \emph{Gaussian processes (GP)} to model single-family houses in Baton Rouge, Louisiana. A frequently used tool to investigate the hypothesis of SVC in an exploratory manner is \emph{geographically weighted regression (GWR)}. For instance, \citet{GWR:Singapore1} and \citet{GWR:singapore2} show that the coefficients of the floor level and the distance to a central business district are spatially varying. We use these findings to motivate a first exploratory analysis of the real estate data set at hand. A visual inspection of the SVCs is challenging since the underlying effects are not directly observable and first require a definition of a regression model. 

\begin{figure}[!ht]
	%% 07.../CV6/GWR_facets
	\centering
	\includegraphics[scale=0.3]{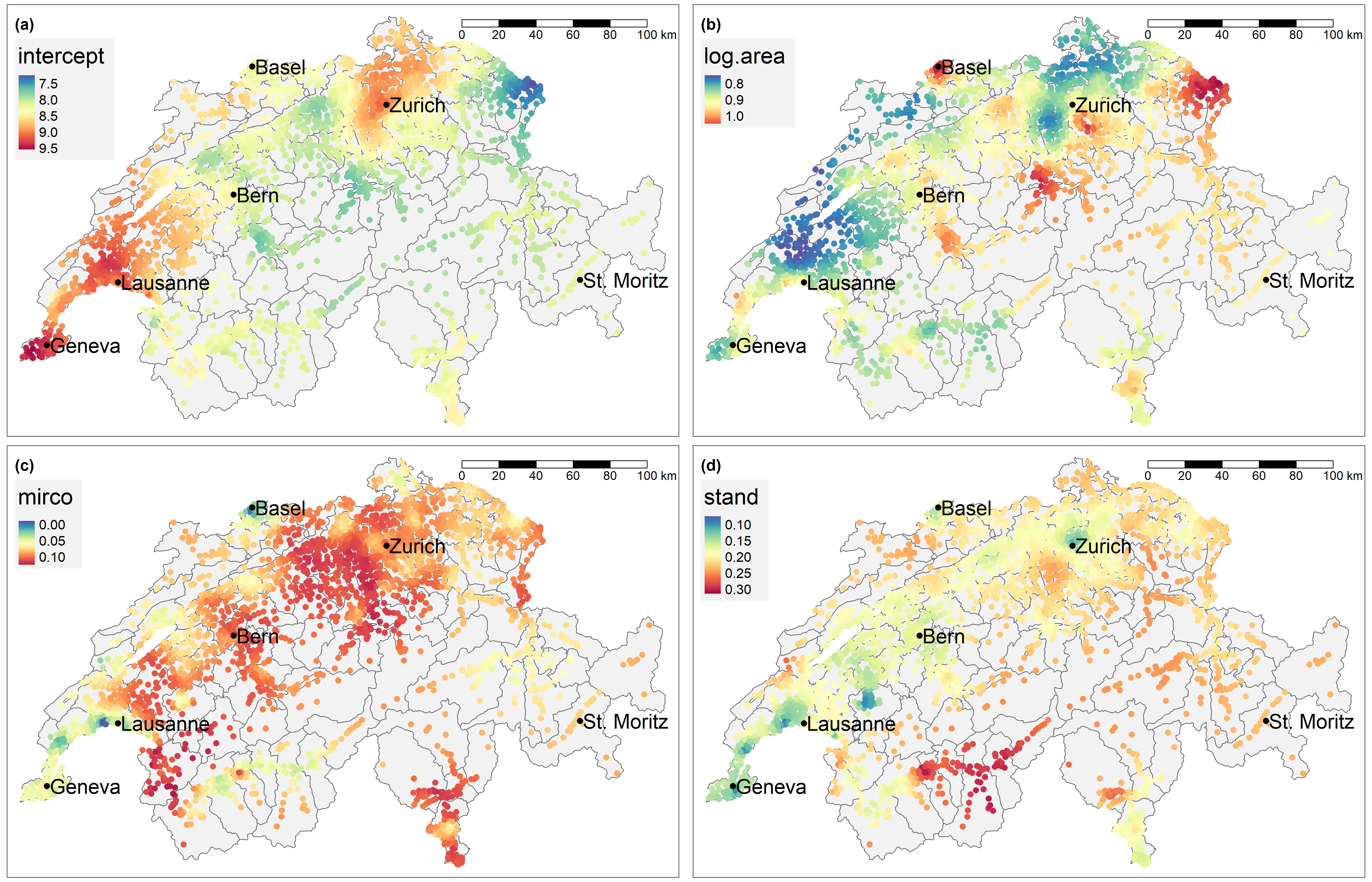}%\includegraphics[scale=0.3]{../figures/use_case/GWR_facet_rev1.png}
	\caption{GWR-estimated SVCs of a model regressing the $\log \texttt{price}$ on (a) an intercept, (b) $\log \texttt{area}$, (c) \texttt{micro}, and (d) \texttt{stand}. \revone{The selected adaptive bandwidth is 222 nearest neighbors.}} \label{fig:GWR_facets}
\end{figure}

With the transaction price (\texttt{price}) as a variable of interest we use the area (\texttt{area}), micro location rating (\texttt{micro}), and standard rating (\texttt{stand}) for a simple, first model. Specifically, we natural logarithm transform the price and area variables as one usually does in a hedonic model \citep{Malpezzi2002, Wheeler2014}. Using the \textsf{R} package \revone{\texttt{GWmodel} by \citet{Gollini2015}}, the model is fitted on the whole data set and the estimated SVCs are depicted in Figure~\ref{fig:GWR_facets}. \revone{Due to the heterogeneous distribution of observation locations, we use an adaptive bandwidth which has been estimated using an automated, AIC corrected selection approach \citep{Hurvich1998}.}

A visual inspection of the GWR-estimated coefficients indicates that we have indeed spatially varying coefficients. In facet (a) the intercept's SVC is given. As expected, we see a relatively large variation in the intercept. \revone{In addition, the covariates do appear to have spatially varying effects.} However, some of the effects are not in line with prior expectation as, for instance, we would expect a higher area effect in city centers. Further, the intercept and the area effect appear negatively correlated. In fact, multicollinearity is a potential drawback of GWR \citep{Wheeler2005, Wheeler2007}.

Yet, applications in context of real estate pricing like \citet{GWR:singapore2} as well as \citet{GWR:Shenzhen} report a substantial increase in $R^2$ by 0.19 and 0.23, respectively, compared to an OLS-based regression. We were able to observe an increase in $R^2$ of similar magnitude when going from an OLS to an GWR estimation. This ambiguity between, on the one hand, the quality of the estimated SVC and, on the other hand, the goodness of fit underlines the need for a statistical sound methodology for SVC models which can be applied to large data. 

\section{SVC Models}\label{sec:SVCmodels}

SVC models extend the linear regression model 
\begin{align}\label{eq:lin_reg}
	y_i = \beta_1 x_i^{(1)} + ... + \beta_p x_i^{(p)} + \varepsilon_i, \quad \varepsilon_i \stackrel{\textnormal{iid}}{\sim} \mathcal{N} (0, \tau^2),
\end{align}
where $i = 1, ..., n$ are the observations and $p$ is the number of coefficients. By allowing the coefficients to vary spatially, the model equation changes to
\begin{align*}
	y_i = \beta_1(s_i) x_i^{(1)} + ... + \beta_p(s_i) x_i^{(p)} + \varepsilon_i, \quad \varepsilon_i \stackrel{\textnormal{iid}}{\sim} \mathcal{N} (0, \tau^2),
\end{align*}
where $s_i \in D \subset \mathbb{R}^d, d \geq 1,$ denotes the location of observation $i$ in a domain $D$. Here, we will work with $d = 2$. The exact specifications for the coefficients $\beta_j(\cdot)$ have yet to be defined.

\subsection{Existing SVC Methods} \label{sec:SVCmethods}% what kind of methods exist?

We will give an overview of the most common methods that allow us to make inference for SVC models.
	
	\paragraph{Geographically Weighted Regression (GWR)} This method is widely used in practice as it is easy to implement and relatively fast in computation. It assumes model equation \eqref{eq:lin_reg} and estimates the coefficients for each location as a weighted regression specific to that location. GWR is fully described in \citet{GWR2002}. As mentioned before, multicollinearity issues with local regressions are raised \citep{Wheeler2005, Wheeler2007}, cf.\ Figure~\ref{fig:GWR_facets}. \revone{This method is readily available for \textsf{R} (e.g.\ packages \texttt{GWmodel} by \citealp{Gollini2015}, \texttt{spgwr} by \citealp{R:spgwr}, and \texttt{gwrr} by \citealp{R:gwrr}) as well as \mbox{arcGIS} \citep{arcgis}. In its classical form GWR only supports the same bandwidth for all coefficients, that is, the bandwidth is either a fixed distance or adaptive defined by a number of nearest neighbors. Recent works by \citet{Fotheringham2017} and \citet{Chen2020} present methods to estimate separate bandwidths for each SVC.} 
	
	\paragraph{Eigenvector Spatial Filtering (ESF)} This method is also known as \emph{Moran's Eigenvector Mapping (MEM)} \citep{Griffith2011, Dray2006}. In \citet{Murakami2015} it has been extended to random effects and thus is now capable of dealing with SVC models. It is readily available in the \revone{\textsf{R} packages \texttt{spmoran} \citep{R:spmoran} and \texttt{spatialreg} \citep{Bivand2015}.}

	\vspace{2em}

The following methods are based on a model assumption where the SVC is defined with Gaussian processes (GP). That is,
\begin{align} \label{eq:beta}
  \beta_j(\cdot) \sim \mathcal{GP} \left(\bm \mu_j (\cdot), c^{(j)}(\ \cdot \ ; \bm \theta_j )\right),
\end{align}
for some choice of covariance function $c^{(j)}$ with parameter vector $\bm \theta_j$. The covariance function works on the distances $r$ between observation locations defined by some norm. Here we use the Euclidean norm $\| \cdot \|$. We denote $\bm \Sigma^{(j)}$ the covariance matrix defined as
\begin{align} \label{eq:covmats}
	\left( \bm \Sigma^{(j)} \right)_{kl} := c^{(j)} (\underbrace{\| s_k - s_l\|}_{= r} ; \bm \theta_j).
\end{align}

One of the most commonly used classes of covariance functions is the Mat\'ern covariance function class. For a \emph{marginal variance} $\sigma^2$ and a \emph{range parameter} $\rho$, we define the Mat\'ern covariance function $c_\nu\left(\ \cdot \ ; \rho, \sigma^2 \right): \left[0, \infty \right) \rightarrow \mathbb{R}^{+}$ as
\begin{align}\label{eq:matern}
		c_\nu \left(r; \rho, \sigma^2 \right) = \sigma^2  \frac{2^{1- \nu} }{\Gamma(\nu)} \left(\sqrt{2\nu} \frac{r}{\rho}\right)^\nu K_\nu \left(\sqrt{2\nu} \frac{r}{\rho}\right),
\end{align} 
where  $\nu \in \mathbb{R}^{+}$ is the \emph{smoothness}, $r$ is a distance, and $K_\nu$ is the modified Bessel function of the second kind and order $\nu$. Coming back to our assumption \eqref{eq:beta}, i.e., that each SVC is modeled by a GP, this implies that each SVC is described by the covariance parameters $\bm \theta_j$ that consists of a variance $\sigma_j^2$ and range $\rho_j$, while in this paper we assume the smoothness $\nu_j$ to be known.

	\paragraph{Bayesian SVC Processes} \citet{SVC2003} introduced a Bayesian SVC model. It allows for prior-dependence of the coefficient processes but assumes an equal range parameter for all coefficients, i.e., $\rho_j \equiv \rho, \forall j$. The method is implemented for example in the \textsf{R} package \texttt{spTDyn} \citep{Bakar2016}. However, the package does not scale to large data.

	\paragraph{Gaussian Markov Random Fields using an SPDE Link} One can define a Bayesian SVC model using the link between Gaussian Markov random fields (GMRF, see \citealp{GMRF:2005}) and GP via a stochastic partial differential equation (SPDE, see \citealp{Lindgren2011}). However, this SPDE link only exists for a limited number of Mat\'ern class covariance functions. Estimation can be done using \emph{integrated nested Laplacian approximations (INLA)} based on \citet{Rue2009} which is available in the \textsf{R} package \texttt{INLA} (\url{www.r-inla.org}, \citealp{R:INLA}). Due to its widely accommodated models, INLA has become quite popular over the last couple of years and is used in environmental sciences and climatology as it can deal with big data sets. A drawback of INLA is the critical assumption on the number of hyperparameters that one can estimate which should be ``small, typically 2 to 5, but not exceeding 20'' \citep{INLA:review2016} and therefore the number of SVCs in a model is limited. 
	
\begin{remark}[SPDE link in INLA]
	In the current version of \texttt{INLA} (version 19.09.03), the SPDE link is defined for the fractional operator order $\alpha \in \left( 0, 2\right]$. In $d$ dimensions, the relation between the fractional operator order $\alpha$ and the Mat\'ern smoothness parameter $\nu$ is $\alpha = \nu + d/2$. In our case ($d=2$) the link exists for Mat\'ern covariance functions with smoothness $\nu \in \left(0, 1\right]$.
\end{remark}

Finally, we want to mention a recent proposal named spatial homogeneity pursuit of regression coefficients by \citet{Li2018}. \emph{Spatially clustered coefficient (SCC)} models are a sub-class of SVC models. While general SVC models usually assume (smooth) spatial variation, SCCs are defined with constant patches and discontinuities in their coefficients. \citet{Li2018} use minimum spanning trees -- a method from graph theory -- to model these SCCs.

\subsection{Challenge}

Our data model (see below in Section~\ref{sec:SVC_RE}) contains $p = 8$ SVCs and about $n = 15{,}000$ observations. While the sample size itself poses no computational problems for existing geostatistical approaches for large data, the combination of the sample size and the number of SVCs is challenging. In particular, when applying existing statistical SVC approaches such as in e.g. \citet{SVC2003} or \citet{Franco2018} to this data, one currently runs into a computational bottleneck. This highlights the need for a statistical methodology that can deal with large data for SVC models.

\section{Method: Maximum Likelihood Estimation for SVC Models} \label{sec:MLE}

In this section, we present the SVC model we use and show how it can be estimated using MLE. We provide novel proposals on how to deal with large data as well as regularization options that help to alleviate correlation among hyper-parameters. Finally, we show how to give predictions once the model has been estimated.

\subsection{Gaussian Process-based SVC Model}\label{sec:notation}

For each covariate $j$ we assume that the associated coefficient $\beta_j(\cdot)$ is separated into a fixed and a random effect. That is, $\beta_j(\cdot) = \mu_j + \eta_j(\cdot)$ for some constant $\mu_j$ and a zero-mean GP $\eta_j(\cdot)$ with a stationary covariance function $c^{(j)}$ similarly to \eqref{eq:beta}, i.e.,
\begin{align} \label{eq:beta0}
  \eta_j(\cdot) \sim \mathcal{GP} \left(\mathbf{0}, c^{(j)}(\ \cdot \ ; \bm \theta_j \  )\right).
\end{align}
We denote by $\Xs \in \mathbb{R}^{n\times p}$ the data matrix defined as $\left(\Xs\right)_{ij} :=x_i^{(j)}$, i.e., the $i$th observation of the $j$th covariate. The fixed effect part is given by $\Xs \bm \mu$, where $\bm \mu := (\mu_1, ..., \mu_p)^\top \in \mathbb{R}^p$. 

Let $\{s_i \}_{i = 1, ..., n}$ be not necessarily distinct observation locations. Using \eqref{eq:covmats} and \eqref{eq:beta0}, $\bm \eta_j := \left(\eta_j(s_1), ..., \eta_j(s_n)\right)^\top \in \mathbb{R}^n$ is normally distributed as
\begin{align*}
  \bm \eta_j \sim \mathcal{N}_n \left(\mathbf{0}_n, \bm \Sigma^{(j)} \right).
\end{align*}
The assumption of mutual prior independence of the GPs, i.e., of $\bm \eta_j$, results in $\bm \eta := (\bm \eta_1, ..., \bm \eta_p)^\top \in \mathbb{R}^{np}$ having the joint distribution $\bm \eta \sim \mathcal{N}_{np} \left(\mathbf{0}_{np}, \bm \Sigma_{\bm \eta} \right)$
with joint covariance matrix
\begin{align} \label{eq:jointcovmat}
	\bm\Sigma_{\bm \eta} := \text{diag}\left( \bm \Sigma^{(1)}, ..., \bm \Sigma^{(p)} \right).
\end{align}

Further, we denote by $\Ws \in \mathbb{R}^{n\times (np)}$ a sparse matrix defined as 
\begin{align*}
  \Ws := \left( \text{diag}(\mathbf{x}^{(1)}) \right| ... \left| \text{diag} (\mathbf{x}^{(p)} )\right). 
\end{align*}
Using this notation, the random effect part is given by $\Ws \bm \eta$. With the identity matrix $\textbf{I}_n$, the error term is distributed as $\epss \sim \mathcal{N}_n \left(\mathbf{0}_{n}, \tau^2 \textbf{I}_n \right)$ and is independent of $\bm \eta$. In summary, writing the response as an $n$-dimensional vector $\mathbf{Y}$, we obtain the GP-based SVC model
\begin{align}\label{eq:SVCmodelmat}
	\mathbf{Y} = \Xs \bm \mu + \Ws \bm \eta + \epss .
\end{align}

\subsection{Likelihood and Optimization}

In the following, we derive the log-likelihood (LL) function for the GP-based SVC model as given in \eqref{eq:SVCmodelmat}. The distribution of the response variable is given by 
\begin{align*}
\yy \sim \mathcal{N}_n \left(\Xs \bm \mu,\bm \Sigma_{\yy} := \Ws \bm\Sigma_{\bm \eta} \Ws^\top + \tau^2 \textbf{I}_n \right).
\end{align*}
Given the observed data, the log-likelihood function depends on the covariance parameters $\bm \theta := (\rho_1, \sigma_1^2, ..., \rho_p, \sigma_p^2, \tau^2)$ as well as the mean parameters $\bm \mu$:
\begin{align*} 
	\text{LL}_{\ys}\left(\bm \theta, \bm \mu \right) &= 
	-\frac{1}{2} \left( \vphantom{\left(\ys - \Xs \bm \mu \right)^\top \bm \Sigma_{\yy}^{-1}}
		n \log (2\pi) + 
		\log \left( \det \bm \Sigma_{\yy} \right) + \left(\ys - \Xs \bm \mu \right)^\top \bm \Sigma_{\yy}^{-1} \left(\ys - \Xs \bm \mu \right) 
	\right).
\end{align*}
Maximizing $\text{LL}_{\ys} \left(\bm \theta, \bm \mu \right)$  is equivalent to minimizing the function
\begin{align*}
	\text{n2LL}_{\ys} \left(\bm \theta, \bm \mu \right) =  
		\log \left( \det \bm \Sigma_{\yy} \right) + 
		\left(\ys - \Xs \bm \mu \right)^\top \bm \Sigma_{\yy}^{-1} \left(\ys - \Xs \bm \mu \right).
\end{align*}
The solution to the optimization problem 
\begin{align*}
\argmin{(\bm \theta, \bm \mu) \in \bm \Omega} \text{n2LL}_{\ys} \left(\bm \theta, \bm \mu \right) 
\end{align*}
with $\bm \Omega := (0, \infty)^{2p+1} \times \mathbb{R}^p \subset  \mathbb{R}^{3p+1}$ cannot be computed analytically and one relies on numerical optimization. We use the quasi Newton method \texttt{"L-BFGS-B"} by \citet{Byrd1995} implemented in the \textsf{R} function \texttt{optim} to numerically minimize $\text{n2LL}_{\ys}$. This optimization approach repeatedly requires the computation of $\bm \Sigma_{\yy}$ with updated parameters $\bm \omega$ and subsequently computing both the determinant and the inverse of $\bm \Sigma_{\yy}$. Both of these tasks are computationally intensive.

\subsubsection{Large Data}

It is at this point that the computational burden of constructing $\bm \Sigma_{\yy}$ and computing its determinant as well as its inverse becomes apparent. Recall that $\Ws$ is a matrix of dimension $n \times (np)$ and $\bm\Sigma_{\bm \eta}$ is of dimension $(np) \times (np)$. Using the formal definition of $\bm \Sigma_{\yy} =  \Ws\bm\Sigma_{\bm \eta} \Ws^\top + \tau^2 \textbf{I}_n$ and the sparsity of $\Ws$, one can verify that the construction of $\bm \Sigma_{\yy}$ alone using the naive matrix multiplication is of run time $\mathcal{O}\left( n^2p^2 \right)$. A Cholesky-decomposition is then being used to compute both the determinant and the inverse of $n \times n$ matrix $\bm \Sigma_{\yy}$ more efficiently. This however has also runtime $\mathcal{O}(n^3)$.

To reduce the computational load, we will exploit the mutual prior independence of the GPs. We introduce the \emph{outer product} of a covariate $\textbf x^{(j)}$ as $\mathbb{X}^{(j)} := \xj \left( \xj \right)^\top$. This allows us to write $\bm \Sigma_{\yy}$ using the Hadamard product (also known as the Schur or direct product) $\odot$ as:
\begin{align} \label{eq:ycovmat}
	\bm \Sigma_{\yy} = \left( \sum_{j = 1}^p \bm \Sigma^{(j)} \odot \mathbb{X}^{(j)} \right) + \tau^2 \textbf{I}_n.
\end{align}

Therefore, we do not have to compute the full matrix multiplication $\Ws\bm\Sigma_{\bm \eta} \Ws^\top$ and the runtime for the construction of \eqref{eq:ycovmat} is $\mathcal{O}(n^2p)$. To reduce the run time for the Cholesky-decomposition, we use covariance tapering proposed by \citet{Furrer2006}. In this approach, we taper the covariance matrices $\bm \Sigma^{(j)}$ by multiplying them with an appropriate compactly supported correlation matrix, say, $\textbf C_{\rho^\star}$, where $\rho^\star$ is the tapering range. Given the underlying covariance functions $c^{(j)}$, without loss of generality, one can choose one corresponding function $c^\star$ which is compactly supported on $\left[0, \rho^\star \right]$ that defines the correlation matrix $\textbf C_{\rho^\star}$ \citep{Furrer2006}. Then the tapered covariance matrix $\bm \Sigma^{(j)}_{\text{tap}} := \bm \Sigma^{(j)} \odot \textbf C_{\rho^\star}$ is sparse with $(\bm \Sigma^{(j)}_{\text{tap}})_{kl} = 0$ for $\| s_k - s_l  \| \geq  \rho^\star$. Using \eqref{eq:ycovmat}, one can easily verify that 
\begin{align*} 
	\bm \Sigma_{\yy, \text{tap}} := \left( \sum_{j = 1}^p \bm \Sigma^{(j)}_{\text{tap}} \odot \mathbb{X}^{(j)} \right) + \tau^2 \textbf{I}_n 
	                              =  \left( \sum_{j = 1}^p \bm \Sigma^{(j)} \odot \mathbb{X}^{(j)} \right) \odot \textbf C_{\rho^\star} + \tau^2 \textbf{I}_n
\end{align*}
is sparse, too.

\subsubsection{ML Estimate Using Direct Optimization Procedure}\label{sec:LL}

Using a straightforward optimization approach the ML estimate is defined as 
\begin{align*}
\widehat{\bm \omega}_{\text{ML}} := \argmin{(\bm \theta, \bm \mu) \in \bm \Omega} \text{n2LL}_{\ys} \left(\bm \theta, \bm \mu \right).
\end{align*} 
When increasing the number of SVCs $p$, the dimension of parameter space $\bm \Omega$ increases and the optimization becomes computationally expensive and numerically unstable. Thus it is crucial to reduce the dimension of the parameter space $\bm \Omega$ when working with many SVC. We solve this problem by proposing to optimize the profile likelihood in $\bm \theta$, which is given by:
\begin{align*}
	\argmin{\bm \theta  \in \bm \Theta} \text{n2LL}_{\ys} \left(\bm \theta, \widehat{\bm \mu}_{\text{GLS}}(\bm \theta) \right),
\end{align*}
where $\bm \Theta := (0, \infty)^{2p+1}$ and $\widehat{\bm \mu}_{\text{GLS}}(\bm \theta)$ is the generalized least squares estimator, i.e.,
\begin{align*} 
	\widehat{\bm \mu}_{\text{GLS}}(\bm \theta) := \left(\Xs^\top \bm \Sigma_{\yy}^{-1}(\bm \theta ) \Xs \right)^{-1} \Xs^\top  \bm \Sigma_{\yy}^{-1}(\bm \theta ) \ys.
\end{align*}
The ML estimate is given by numerically optimizing the following:
\begin{align*}
	\widehat{\bm \omega}_\text{ML} &:= (\widehat{\bm \theta}_\text{ML}, \widehat{\bm \mu}_\text{ML}), \quad \text{where}   \\ 
	\widehat{\bm \theta}_\text{ML} &:= \argmin{\bm \theta  \in \bm \Theta} \text{n2LL}_{\ys} \left(\bm \theta, \widehat{\bm \mu}_{\text{GLS}}(\bm \theta) \right), \\
	\widehat{\bm \mu}_{\text{ML}}   &:= \widehat{\bm \mu}_{\text{GLS}}(\widehat{\bm \theta}_\text{ML}). 
\end{align*}

\subsubsection{Regularization Using PC Priors}\label{sec:regLL}

Due to weak identifiability and posterior correlation, the optimization concerning the covariance parameters can be unstable. We want to apply some form of regularization to ensure the numerical optimization problem is well-posed. Recent advances by \citet{Simpson2017} and extensions thereof by \citet{Fulgstad2018} introduced penalizing complexity (PC) priors  for Gaussian random fields of Mat\'ern class. We use these PC priors as regularizers to construct a regularized likelihood by extending our pure likelihood or profile likelihood approach from Section~\ref{sec:LL}, respectively. In the following, we will define the \emph{regularized parameter estimate}. 

In our parametrization of the Mat\'ern covariance function and with $d = 2$, the PC priors for a single GP with range $\rho$ and marginal standard deviation $\sigma$ \citep[Theorem~2.6]{Fulgstad2018} take the form
\begin{align*}
	\pi_\text{PC} (\rho, \sigma) := \lambda_\rho \lambda_\sigma (2\rho)^{-2} \exp \left( -\lambda_\rho (2\rho)^{-1} - \lambda_\sigma \sigma \right),
\end{align*}
where $\lambda_\rho$ and $\lambda_\sigma$ are defined by the prior beliefs on the lower tail of the range, $\mathbb{P}(\rho < \rho_0) = \alpha_\rho$, and the upper tail of the standard deviation, $\mathbb{P}(\sigma > \sigma_0) = \alpha_\sigma$, respectively. They are given by $\lambda_\rho = -2\log(\alpha_\rho) \rho_0$ and $\lambda_\sigma = - \log(\alpha_\sigma)/\sigma_0$. Under iid assumption on each prior and some initial beliefs this defines the regularized estimate using $\text{n2LL}_{\ys}$ as:
\begin{align*}
	\widehat{\bm \omega}_{\text{reg}} := \argmin{(\bm \theta, \bm \mu) \in \bm \Omega} \left( \text{n2LL}_{\ys} \left( \bm \theta, \bm \mu \right) + \sum_{j = 1}^p \left( \frac{\lambda_{\rho_j}}{\rho_j} + 4 \log \rho_j + 2 \lambda_{\sigma_j} \sigma_j \right)\right).
\end{align*}
\subsection{Prediction of Coefficient Processes}

In the following, we describe how to predict the covariate effects at locations that possibly have not been observed, given estimated parameters $\widehat{\bm \omega}_\text{ML}$. In the classical case of predicting a single GP at spatial points the \emph{empirical best linear unbiased predictor (EBLUP)} is used \citep{Cressie1990}. In the case of SVCs, we likewise establish the EBLUP in the following.

We first extend our notation from observed locations $\mathbf{s} = (\ston)^\top$ to the $n'$ locations we want to make predictions for, namely $\mathbf{s}' = (\sptonp)^\top$. These may or may not include already observed locations. The  distributions of $\bm \beta' :=\bm \beta\!\left( \mathbf{s}' \right) = \bm\mu_{\bm \beta'} + \bm \eta\!\left( \mathbf{s}' \right)$ and $\yy$ are again normally distributed with respective means $\bm\mu_{\bm \beta'} := \bm \mu \otimes  \mathbf{1}_{n'}$ and $\textbf{X} \bm\mu_{\betas}$. Estimating the latent coefficient processes is done in a two step approach where we first estimate $\bm \eta' := \bm \eta\!\left( \mathbf{s}' \right)$ and then add the mean estimate of $\bm\mu_{\bm \beta'}$. We start by considering the joint distribution 
\begin{align*}
	\begin{pmatrix}
		\bm \eta' \\
		\yy
	\end{pmatrix} \sim 
	\mathcal{N}_{n'p+n} \left(
		\begin{pmatrix}
			\mathbf{0}_{n'p} \\
			\bm\mu_{\yy}
		\end{pmatrix}, 
		\begin{pmatrix}
			\bm\Sigma_{\bm \eta'} & \bm\Sigma_{\bm \eta' \yy} \\
			\bm\Sigma_{\yy \bm \eta'} & \bm\Sigma_{\yy}
		\end{pmatrix}
	\right).
\end{align*}
The covariance matrix $\bm\Sigma_{\bm \eta'}$ is defined for locations $\sptonp$ in an analogous way to \eqref{eq:covmats} and \eqref{eq:jointcovmat}, namely $\left( \bm \Sigma'^{(j)} \right)_{kl} := c^{(j)}(\|s_k' - s_l' \|; \bm \theta_j)$ and hence 
\begin{align*}
\bm\Sigma_{\bm \eta'}:= \text{diag}\left( \bm \Sigma'^{(1)}, ..., \bm \Sigma'^{(p)} \right). 
\end{align*}
The covariance matrix $\bm\Sigma_{\yy}$ is given in \eqref{eq:ycovmat}. The cross-covariances matrices $\bm\Sigma_{\bm \eta' \yy}$ and $\bm\Sigma_{\yy \bm \eta'}$ are defined as 
\begin{align*}
	\bm\Sigma_{\yy \bm \eta'} &:= \text{Cov}(\Xs \bm \mu + \Ws \bm \eta + \epss, \bm \eta') = \Ws \text{Cov}(\bm \eta, \bm \eta'),\\
	\bm\Sigma_{\bm \eta' \yy} &:= \bm\Sigma_{\yy \bm \eta'}^\top,
\end{align*}
where $\text{Cov}(\bm \eta, \bm \eta')$ is again defined as \eqref{eq:covmats} and \eqref{eq:jointcovmat}, but now with corresponding locations $\ston$ and $\sptonp$. Using the conditional distribution $\bm \eta' | \yy = \ys$ and plugging in $\widehat{\bm \omega}_{\text{ML}}$ one receives the EBLUP for SVC as 
\begin{align*}
	 \widehat{\bm \eta}' := \widehat{\bm\Sigma}_{\bm \eta'  \yy}  \widehat{\bm\Sigma}_{\yy}^{-1}\left(\ys - \widehat{\bm\mu}_{\yy} \right)
\end{align*}
and therefore $\widehat{\bm \beta}' := \widehat{\bm\mu}_{\bm \beta'} + \widehat{\bm \eta}'$. One can then use corresponding data $\Xs'$ and $\Ws'$ at locations $\mathbf{s}'$ to get predictions for $\mathbf{Y}'$. Predictive variances of such $\widehat{\mathbf{Y}}'$ are derived in a similar way as above and given by the diagonal of $\bm\Sigma_{\yy'} - \bm\Sigma_{\yy' \yy} \bm\Sigma_{\yy}^{-1} \bm\Sigma_{\yy \yy'}$.

\subsection{\textsf{R} package \texttt{varycoef}}

The MLE described in this section is implemented in the \revone{\textsf{R} package \texttt{varycoef} \citep{R:varycoef}}. It \revone{utilizes parallel computing as well as sparse matrix representation and computation for numeric optimization procedures \citep{FG2019,R:spam}. This package} is used throughout this work whenever the MLE approach is mentioned. We indicate the usage of our method and the \textsf{R} package \texttt{varycoef} on GP-based \revone{SVC models by the abbreviation MLE with suffixes}.

\section{Simulation Study} \label{sec:sim_study}

This and the next section are designated to compare existing and our proposed methods in regard to parameter estimation and prediction accuracy. From Section~\ref{sec:SVCmethods}, we exclude the Bayesian SVC processes method since it does not scale to large data. Further, since SCC and smooth SVC models are inherently different, we exclude the spatial homogeneity pursuit approach, too.

\subsection{Setup}

In order to empirically validate our method and to compare it to existing methods, we define the following simulation setup. We simulate $N:=100$ times a GP with varying numbers of SVCs $p$ and sample sizes $n'$. Latter is defined by a positive integer $q$ such that $n' = (2q)^2$ is the number of data points and locations which are sampled from a \emph{perturbed grid} \citep{Furrer2016}.

\begin{figure}[!h]
	\centering
	\includegraphics[scale=1]{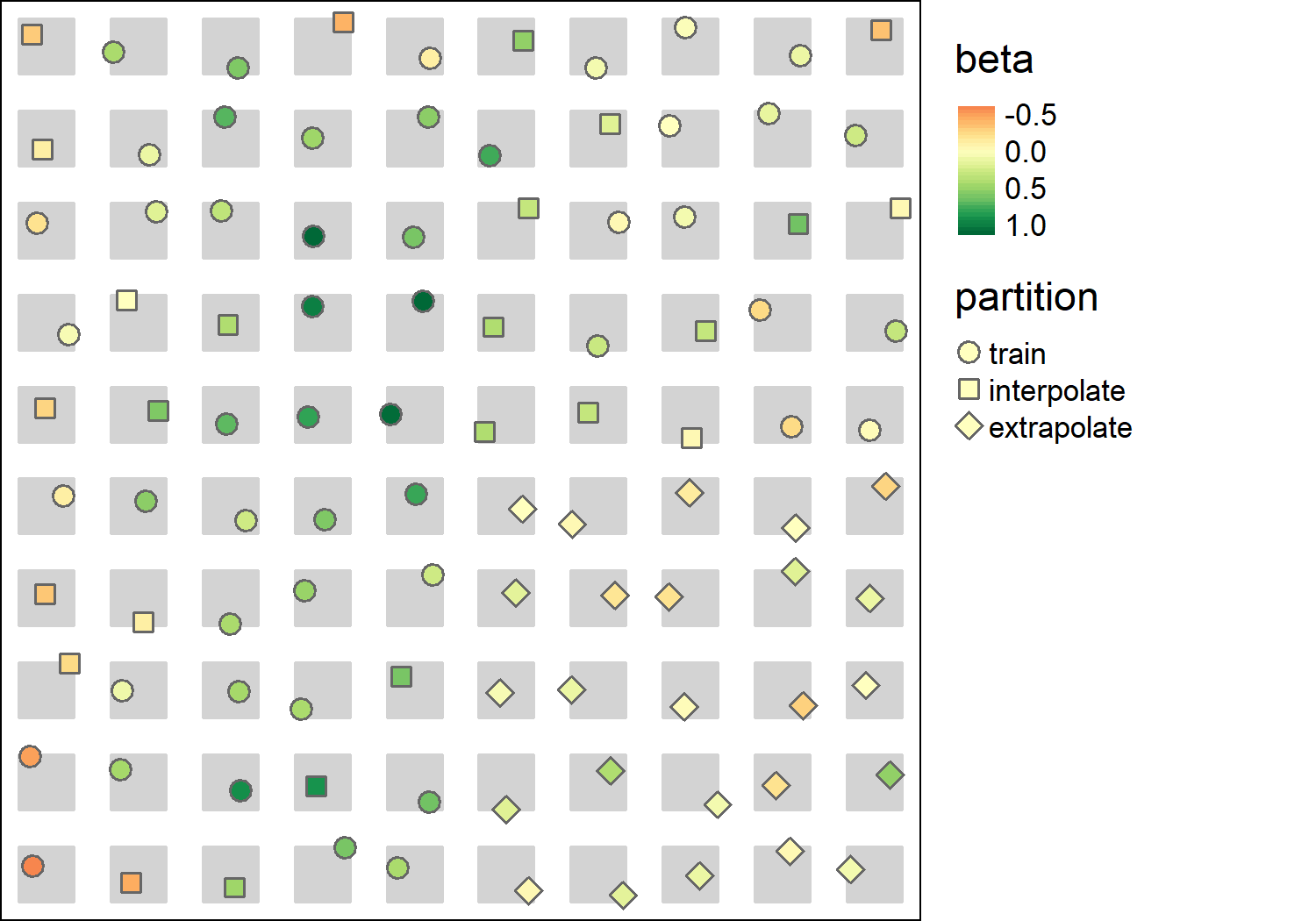}%\includegraphics[scale=1]{../figures/fig_sampling.png}
	\caption{Example of a perturbed grid defined with $q=5$ for $n' = 100$ sample locations and a partition into training, interpolation, and extrapolation testing. The gray patches are the sampling domain. The distance between each patch is at least $\delta/q$. Here and in all of our simulations, $\delta$ is set to $0.2$.} \label{fig:sampling}
\end{figure}

A perturbed grid consists of $(2q) \times (2q)$ unit squares. For each ${(r, s)} \in \{ 0, ...,{2q-1} \}^2$, we uniformly draw a single location from a square ${[r + \delta, (r+1)-\delta]} \times {[s + \delta, (s+1)-\delta]}$, where $\delta \in {[0, 0.5)}$ restricts a unit square area by an outer margin. Finally, we standardize the locations by $(2q)^{-1}$ such that the total domain of a perturbed grid is contained in the unit square. Thus, we receive the sample locations $s_1, ..., s_{n'}$. An example is given in Figure~\ref{fig:sampling}. 

At these locations the SVCs, the error term $\bm \varepsilon$, and the data $\Xs$ are sampled and we compute the response $\ys$. We set $\textbf{x}^{(1)} = \textbf{1}_n$ which allows us to model a spatially varying intercept. The remaining data of $\Xs$ is sampled from a standard-normal distribution for coefficients $j = 2, ..., p$.

The data is then divided into three disjoint folds, \emph{(i)} a training data set $\mathcal{S}_\text{train}$, \emph{(ii)} a test data set for interpolation $\mathcal{S}_\text{interpolate}$, and \emph{(iii)} a test data set for extrapolation $\mathcal{S}_\text{extrapolate}$. The unit square is partitioned into four quadrants. The lower right quadrant is an \emph{extrapolation test set} and contains $25\%$ of the data. In the other quadrants, $25\%$ of the data is randomly assigned as \emph{interpolation test set}. On the rest ($50\%$) of the data, the model is being estimated. Thus, we have a partition: 
	\begin{align*}
		\mathcal{S} &:=\ \mathcal{S}_\text{train}\ \cup\ \mathcal{S}_\text{interpolate}\ \cup\ \mathcal{S}_\text{extrapolate} = \{1, ..., n'\}, \\
		n &:= | \mathcal{S}_\text{train} | =  \frac{n'}{2}, \quad  \quad | \mathcal{S}_\text{interpolate} | =  | \mathcal{S}_\text{extrapolate} | =  \frac{n'}{4}.
	\end{align*}		

In total, there are three simulation settings which differ in the number of total sampled points $n'$ and SVCs $p$ and five methods for SVC modeling, see Table~\ref{tab:sim} for an overview. \revone{For our proposed methodology, we maximize the regularized profile likelihood and label it as MLE.r}. The SPDE method has been implemented with the \textsf{R} package \texttt{INLA} \citep{R:INLA} and uses the same PC priors as MLE.r, namely ${\mathbb{P}(\rho < 0.075)} = 0.05$ and ${\mathbb{P}(\sigma > 0.25)} = 0.05$. \revone{The methods ESF and GWR are implemented with the \textsf{R} packages \texttt{spmoran} \citep{R:spmoran} and \texttt{GWmodel} \citep{Gollini2015}, respectively. For GWR, we use the same bandwidth for all covariates as the prediction function does not support covariate specific bandwidths. The bandwidths are estimated using a cross validation and are not adaptive, i.e., they are defined as fixed distances due to the regular structure of the perturbed grid.} The superscript \emph{tap} in Table~\ref{tab:sim} indicates covariance tapering for MLE, since Simulation~2 has the most observations. The taper range $\rho^\star$ is $0.2$.
Due to the number of SVCs, the SPDE method cannot be applied on Simulation 2, hence the cross marks in Table~\ref{tab:sim}.

\begin{table}[!ht]
\centering
\caption{Overview of simulations and methods.}\label{tab:sim}
\begin{tabular}{{l}{l}|*{3}{c}}
	\hline
	 			&		& \multicolumn{3}{c}{Simulations}   \\
	 			&		& 1 			& 2 				& 3 \\ \hline
			    & $n'$  & $2{,}500$		& $10{,}000$		& $2{,}500$ \\
			    & $n $  & $1{,}250$		& $5{,}000$			& $1{,}250$ \\
			    & $p$   & 3 			& 3					& 10 \\ \hline
	Methods		&		&				&					& \\ \hline
	MLE.r       &       & $\checkmark$	& $\checkmark^{tap}$& $\checkmark$ \\
	SPDE        &	    & $\checkmark$	& $\checkmark$		& \ding{55} \\
	ESF			&	    & $\checkmark$	& $\checkmark$		& $\checkmark$ \\
	GWR			&	    & $\checkmark$	& $\checkmark$		& $\checkmark$ \\
	\hline           
\end{tabular}
\end{table}

In each repetition $w = 1, ..., N$ and for each method $m$, we calculate the RMSE between the estimated SVC $\widehat{\bm \beta}_j^{(m, w)}$ and the true SVC $\bm \beta_j$ in each fold 
\begin{align*}
	\kappa \in \{\text{train}, \text{interpolate}, \text{extrapolate}\}.
\end{align*}
Analogously, we calculate the RMSE for the response. Thus, we have:
\begin{align}
	\text{RMSE}_{\kappa}^{(m, w)} \left( \bm \beta_j \right) &= \sqrt{\frac{1}{\left|\mathcal{S}_\kappa\right|} \sum_{i \in \mathcal{S}_\kappa }\left| \beta_j(s_i) - \widehat{\beta}_j^{(m, w)} (s_i) \right|^2},\label{eq:RMSEbeta}\\
	\text{RMSE}_{\kappa}^{(m, w)} \left( \textbf y \right) &= \sqrt{\frac{1}{\left|\mathcal{S}_\kappa\right|} \sum_{i \in \mathcal{S}_\kappa }\left| y(s_i) - \widehat{y}^{(m, w)} (s_i) \right|^2}.\label{eq:RMSEy}
\end{align}
 
In all of our simulation studies we assume the type of covariance function to be known, which is why we have to define it here. Since we expect in most of our applications the fields to be not too smooth, we follow the recommendation of \citet{SteinMichaelL1999IoSD} and use exponential covariance functions. The exponential covariance function is a special case of the Mat\'ern class covariance functions with smoothness parameter $\nu = 1/2$, cf.~\eqref{eq:matern}. Thus, we have
\begin{align*}
	\left(\bm \Sigma^{(j)} \right)_{kl} := c_{1/2} \left( \|s_k - s_l \| ; \bm \theta_j \right) = \sigma^2_j \exp\left( -\frac{\|s_k - s_l\|}{\rho_j}\right), \quad \text{for all } j.
\end{align*}

\subsection{Results}
\subsubsection{Simulation 1: Base Setup}\label{sec:base_study}

\begin{figure}[!ht]
	\centering
	\includegraphics[scale=0.6]{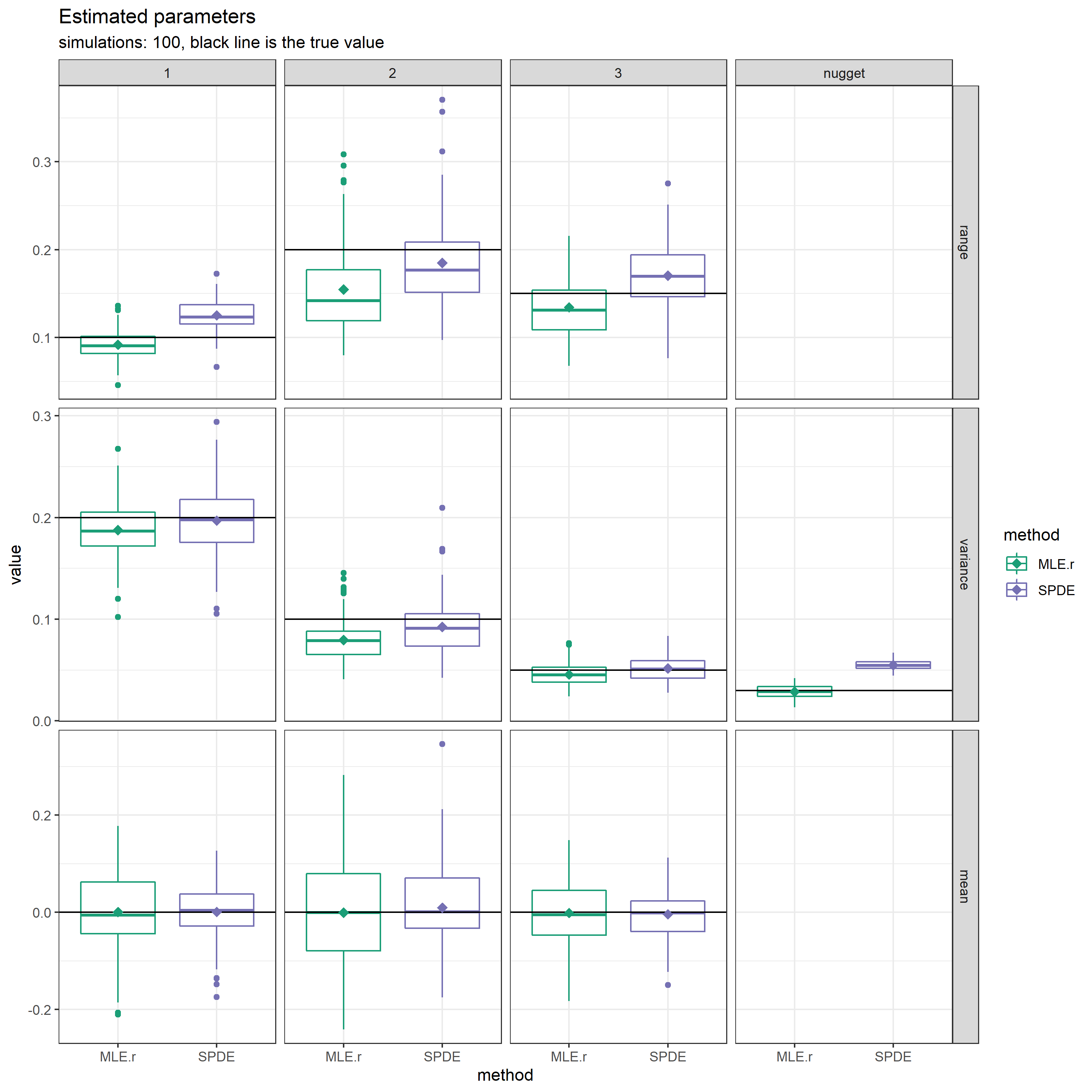}%\includegraphics[scale=0.6]{../code/simulation/plots/sim1_parameters_revision.png}
	\caption{Estimated covariance and mean parameters of MLE and SPDE methods of Simulation~1. The box plots consists of 100 estimations of different realization. The diamonds indicate the means. True values are given by black lines.} \label{fig:sim1_parameters}
\end{figure}

The base simulation setup samples $n' = 2{,}500$ from a perturbed grid with $p = 3$ SVCs. The parameters of the true model are provided in Table~\ref{tab:parameters1}. The results of the parameter estimation which is available \revone{for the  methods MLE.r and SPDE are given in Figure~\ref{fig:sim1_parameters}. The results are quite similar. While the SPDE's estimates of the range usually are higher than those of MLE.r, SPDE overestimates the nugget variance. Regarding the mean effects, SPDE appears to estimate mean effects with more precision.}

\begin{table}[!ht]
\centering
\caption{Parameters of the underlying true model in Simulations 1 and 2.}\label{tab:parameters1}
\setcellgapes{2pt}\makegapedcells
\begin{tabular}{l | *{3}{c}  }
	\hline
	               & \multicolumn{3}{c}{Effects $j$}\\
	Parameters     & 1     & 2      & 3  \\ \hline
	$\mu_j$      & $0.00$ & $0.00$ & $0.00$ \\
	$\rho_j$     & $0.10$ & $0.20$  & $0.15$ \\
	$\sigma^2_j$ & $0.20$ & $0.10$  & $0.05$ \\ \hline
	$\sigma_\text{nugget}^2$ & \multicolumn{3}{c}{$0.03$}\\\hline
	
\end{tabular}
\end{table}

\begin{figure}[!ht]
	\centering
	\includegraphics[scale=0.6]{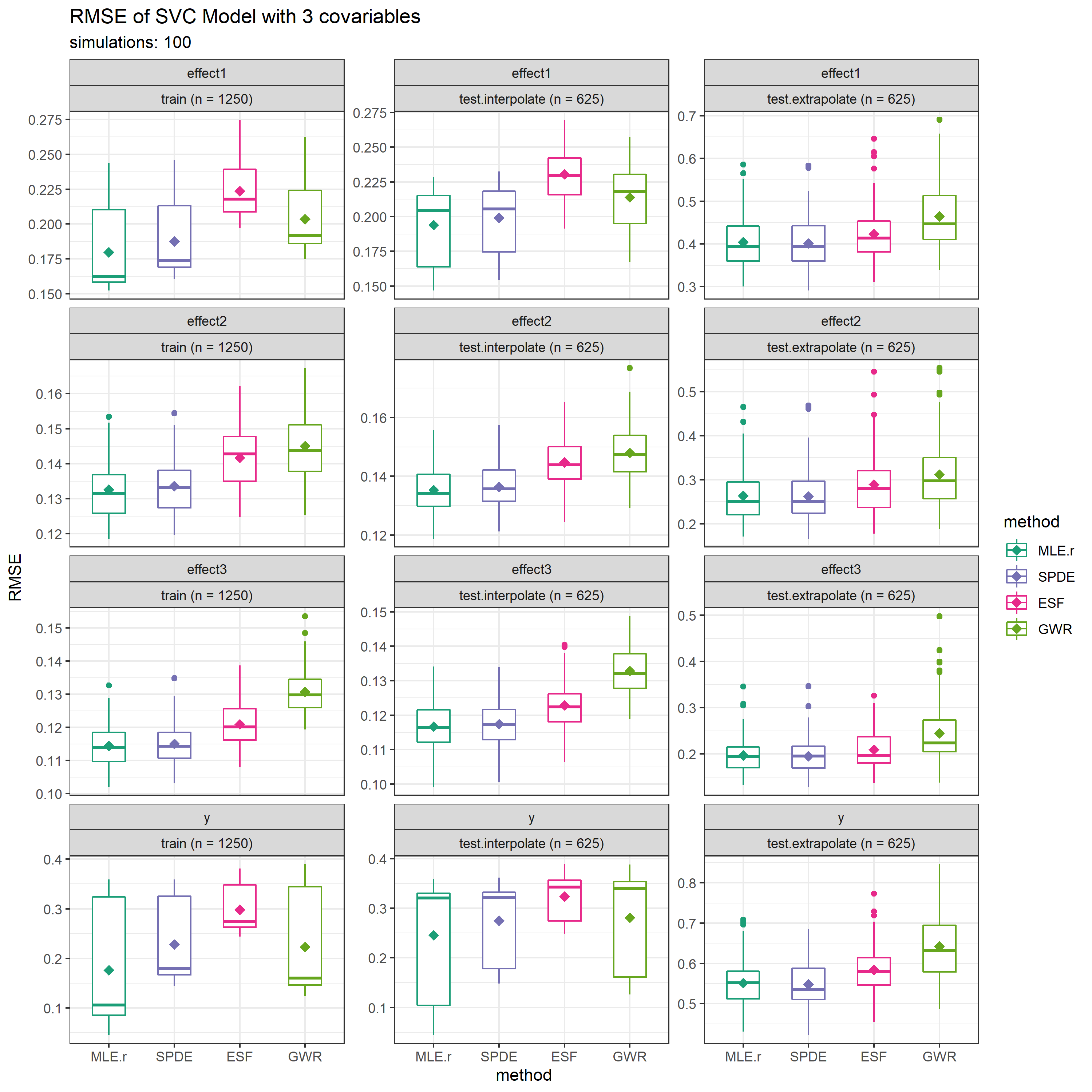}%\includegraphics[scale=0.6]{../code/simulation/plots/sim1_RMSE_all_revision.png}
	\caption{Box plots of RMSE between true $\bm \beta_j$ and $\widehat{\bm \beta}_j$ as well as between $\mathbf y$ and $\mathbf{\widehat{y}}$ for Simulation~1 in corresponding partition, see \eqref{eq:RMSEbeta} and \eqref{eq:RMSEy}. The diamonds indicate the means. Note the different ranges of the $y$-axis.} \label{fig:sim1_RMSE}
\end{figure}

\revone{The results for the RMSEs are depicted as box plots in Figure~\ref{fig:sim1_RMSE} for all 100 repetitions within the simulation. The RMSE as given in \eqref{eq:RMSEbeta} reveals some interesting insights. MLE.r consistently gives some of the best training and prediction results. It is closely followed by SPDE. ESF and GWR clearly have lower accuracy. Spatial extrapolation as depicted in the last column is much more difficult for all methods. But, model-based approaches perform better than the non-model based alternatives. The results for SVC's RMSE probably translate to the results for the response's RMSE, where it appears that the intercept is the main driver. Overall, MLE.r has the highest in-sample and out-of-sample predictive accuracy, closely followed by SPDE. ESF and GWR have lower predictive accuracy.}

\subsubsection{Simulation 2: Number of Observations is \texorpdfstring{$n' = 10{,}000$}{n10000}}

\begin{figure}[!ht]
	\centering
	\includegraphics[scale=0.6]{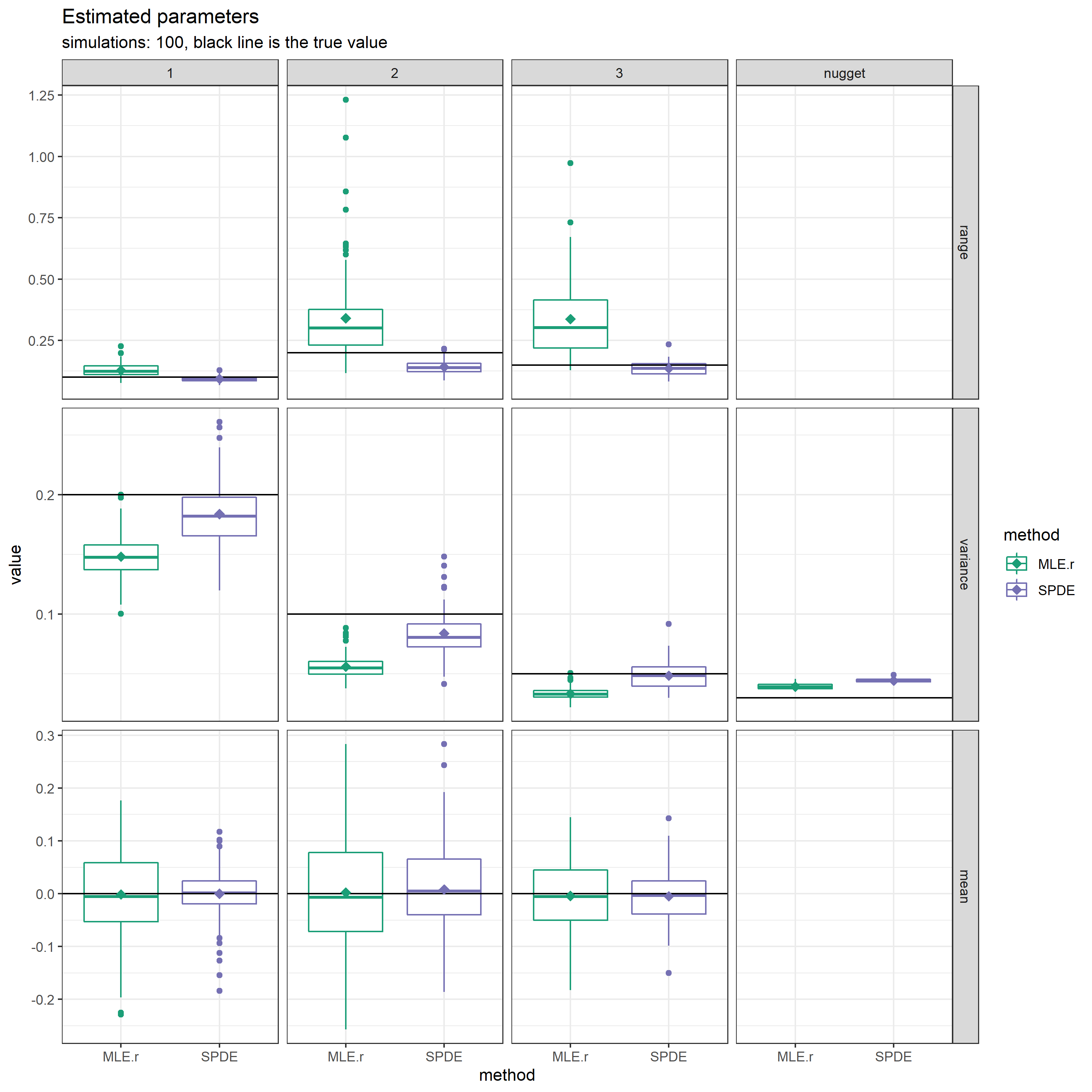}%\includegraphics[scale=0.6]{../code/simulation/plots/sim2_parameters_revision.png}
	\caption{Box plots represent estimated covariance and mean parameters by MLE.r and SPDE methods in Simulation~2. The diamonds indicate the means. \revone{True values are given by black lines.}} \label{fig:sim2_parameters}
\end{figure}

Simulation 2 has the same underlying true model to simulate the data from, cf.~Table~\ref{tab:parameters1}, but using a data set with $10{,}000$ observations. In order to allow our method to scale to data sets with a large $n$, we introduced covariance tapering \citep{Furrer2006} in both MLE without and with regularization. With the introduction of a tapering range, which in this simulation study was set to $\rho^\star = 0.2$, the covariance matrices become sparse. While this has a positive impact on computation time, it results in biased estimates for the covariance parameters. Due to tapering, the ranges were overestimated \revone{by MLE.r}, whilst all variances of the SVCs were underestimated. \revone{SPDE performed very similarly as in Simulation~1, cf.~Figure~\ref{fig:sim2_parameters}.}

\revone{Further, we compute the RMSE as given in \eqref{eq:RMSEbeta} and \eqref{eq:RMSEy}. The results are depicted in Appendix~\ref{appendix:sim2}. They are very similar to what we observed in Simulation~1 with Figure~\ref{fig:sim1_RMSE}. The main difference that we can see is that the GWR has a broad range of quality when estimating and predicting the response, while clearly falling behind when modeling the SVCs.}

\subsubsection{Simulation 3: Number of SVCs is \texorpdfstring{$p = 10$}{p10}}

In this simulation, we have $p = 10$ SVCs sampled from the model with the true parameters as defined in Table~\ref{tab:parameters3}. We cannot run SPDE due to too many SVCs. Thus, \revone{MLE.r is the only method with which we are able to estimate the parameters. The empirical results show that our ML-estimator yields unbiased results, which qualitatively are close to the results of MLE given in Figure~\ref{fig:sim1_parameters}. In predictive performance measured by the RMSE, MLE.r surpasses both ESF and GWR. All of the results are depicted in Appendix~\ref{appendix:sim3}.}

\begin{table}[!ht]
	\centering
	\caption{Parameters of the underlying true model in Simulation 3. Compared to Simulations 1 and 2 the first SVCs are exactly defined as before. The nugget is now larger to ensure similar noise to signal ratio.}\label{tab:parameters3}
	\setcellgapes{2pt}\makegapedcells
	\begin{tabular}{l | *{10}{c} }
	\hline
	             & \multicolumn{10}{c}{Effects $j$} \\
	Parameters   & 1      & 2      & 3      & 4      & 5      & 6      & 7      & 8      & 9      & 10 \\\hline
	$\mu_j$      & $0.00$ & $0.00$ & $0.00$ & $0.00$ & $0.00$ & $0.00$ & $0.00$ & $0.00$ & $0.00$ & $0.00$ \\
	$\rho_j$     & $0.10$ & $0.20$ & $0.15$ & $0.10$ & $0.05$ & $0.05$ & $0.15$ & $0.15$ & $0.20$ & $0.20$\\
	$\sigma^2_j$ & $0.20$ & $0.10$ & $0.05$ & $0.05$ & $0.10$ & $0.05$ & $0.10$ & $0.15$ & $0.15$ & $0.20$\\ \hline
	$\sigma_\text{nugget}^2$ &  \multicolumn{10}{c}{$0.10$} \\
	\hline	
	\end{tabular}
\end{table}

\subsubsection{Summary}

\revone{Over all simulations, our proposed MLE approach is competitive with other methodologies for SVC modeling. In particular, the parameter estimation is unbiased when not using tapering and scales to models with a moderate number of SVCs. Further, the modeling and predicting capabilities as measured using the RMSE are among the best.}

\revone{In the following, we also report computation time. The simulations were run on a server with 8 Intel Xeon 10 core E7-2850 with 2.0~GHz for a total of 80 threads and 2 TB of memory under Ubuntu OS. Computations run time was measured for each repetition $w$ and method $m$ run. The results are given in Table~\ref{tab:time}.}

\begin{table}[!ht]
	\centering
	\caption{\revone{Median time for estimation and prediction out of 100 repetitions given in hours and minutes (${h:mm}$). Note that MLE.r was run on a 79 thread cluster established prior to the simulation. The overhead of initializing the cluster is \emph{not} accounted for, but takes less than a minute. SPDE computations using \texttt{INLA} parallelize on their own. GWR computations include automated fixed bandwidth selections using an CV approach. Further, this is the only method that does the estimation and prediction step at once.}}\label{tab:time}
\begin{tabular}{{l}|*{5}{r}}
	\hline
	 			& \multicolumn{4}{c}{Methods}   \\
	Simulations	& MLE.r	& SPDE & ESF	 	& GWR \\ \hline
	1         	& $0:09$ & $0:10$   & $< 0:01$  & $< 0:01$\\
	2			& $3:37$ & $1:11$	& $0:13$  & $0:02$\\
	3			& $0:16$ & 	& $0:04$  & $< 0:01$\\
	\hline           
\end{tabular}
\end{table}

\revone{Overall, the results are in line with our expectations. Model-based estimation and prediction approaches are much slower than non-model-based procedures. While increasing the number of SVCs $p$, the computation time only increases slightly for all methods, the increase of computation time is much more pronounced when increasing the number of observations. Note that the computational time for MLE.r in the large $n$ case depends on the amount of tapering. Using a smaller taper range would decrease the computational time.}

\section{Application: Real Estate Pricing} \label{sec:CHap}

In prior consultation with real estate experts at Fahrl\"ander Partner, real estate mass appraisal will be done on a model fitted on transactions from six consecutive quarters using (transformed or scaled) covariates given in Table~\ref{tab:covariates}. Predictions are then given for the following seventh quarter. The rationale behind this setup is to account for a time trend.

The focus of this application lies -- like in previous simulation studies -- on two aspects. First, we will compare and analyze the outputs of ML-estimated GP-based models with respect to parameter estimation and interpretation of the estimated SVCs (Section~\ref{sec:mod_inter}). Second, we will expand this frame work in order to compare the predictive performance of the different methods (Section~\ref{subsec:pred_perf}). We use a \emph{moving window} validation. That is, we divide the data set into 4 folds ($f = 1, ..., 4$). Each fold $f$ consists of data from 6 consecutive quarters $\mathcal{S}_{\text{train}, f}$ which are used to estimate the model and a following seventh quarter $\mathcal{S}_f$ which is used for evaluating the predictive accuracy. A visualization of the moving windows is given in Table~\ref{tab:moving_window}. 

\begin{table}[!ht]
	\centering
	\caption{Moving window validation set up with four folds. The model is fitted on all observations with $s_i \in \mathcal{S}_{\text{train}, f}$ and tested out-of-sample on $\mathcal{S}_f$.} \label{tab:moving_window}
	\begin{tabular}{{l} | *{10}{c}}
	\hline
			 & \multicolumn{2}{c}{2015} & \multicolumn{4}{c}{2016} & \multicolumn{4}{c}{2017} \\
	Folds $f$ & Q3 & Q4 & Q1 & Q2 & Q3 & Q4 & Q1 & Q2 & Q3 & Q4 \\ \hline
	1 & \multicolumn{6}{c}{\cellcolor{Gray}$\mathcal{S}_{\text{train}, 1}$} & \cellcolor{lightGray} $\mathcal{S}_1$&  \\
	2 & & \multicolumn{6}{c}{\cellcolor{Gray}$\mathcal{S}_{\text{train}, 2}$} & \cellcolor{lightGray} $\mathcal{S}_2$ & \\
	3 & & & \multicolumn{6}{c}{\cellcolor{Gray}$\mathcal{S}_{\text{train}, 3}$} & \cellcolor{lightGray} $\mathcal{S}_3$ & \\
	4 & & & & \multicolumn{6}{c}{\cellcolor{Gray}$\mathcal{S}_{\text{train}, 4}$} & \cellcolor{lightGray} $\mathcal{S}_4$ \\
	 \hline
\end{tabular}
\end{table}

\subsection{SVC Model for Real Estate Pricing} \label{sec:SVC_RE}

We extend the previous simple model used in Section~\ref{sec:motivation} and add \emph{i)} a standardized age $\texttt{Z.age} = (2000- \texttt{yoc})/20$ defined with the year of construction \texttt{yoc}, \emph{ii)} $\texttt{Z.age}^2$ as one expects a quadratic age effect in hedonic models in Europe \citep{ViennaAMM2010, SF2006}, \emph{iii)} the renovation rating \texttt{renov}, and \emph{iv)} a dummy variable constructed from the quarter of transaction. Latter is defined as
\begin{align*}
	\texttt{D}_\text{lastQ} := \begin{cases}
	1, & \text{if transaction took place in the last quarter of the training data,} \\
	0, & \text{otherwise.}
	\end{cases}
\end{align*}
This is to differentiate the most recent transactions from the rest of the training set in order to account for the temporal trend and should enhance the predictive performance. In summary, we will obtain the following model with $p = 8$ SVCs:
\begin{equation}\label{eq:SVCmodelCHap}
\begin{aligned}
	y_i := \log\texttt{price}_i &= \quad \beta_1(s_i)   &+\ &\beta_2(s_i)\ \log \texttt{area}_i    &+\ &\beta_3(s_i)\ \texttt{Z.age}_i \\ 
			  &\quad+ \beta_4(s_i)\ \texttt{Z.age}_i^2 &+\ &\beta_5(s_i)\ \texttt{micro}_i        &+\ &\beta_6(s_i)\ \texttt{stand}_i \\
	     	  &\quad+ \beta_7(s_i)\ \texttt{renov}_i   &+\ &\beta_8(s_i)\ \texttt{D}_\text{lastQ} &+\ &\varepsilon_i, 
\end{aligned}
\end{equation}
where the locations $s_i$ are given in \emph{transformed} LV03 coordinates. More precisely, the easting (\texttt{LV03x}) and northing (\texttt{LV03y}) are now centered on the origin $(0; 0)$ and transformed to kilometers, i.e.
\begin{align*}
	\begin{pmatrix}
		\texttt{Z.LV03x} \\
		\texttt{Z.LV03y}
	\end{pmatrix} := 10^{-3}\cdot \left( \begin{pmatrix}
		\texttt{LV03x} \\
		\texttt{LV03y}
	\end{pmatrix} - \begin{pmatrix}
		600{,}000 \\
		200{,}000
	\end{pmatrix} \right) .
\end{align*}
This procedure increases numerical stability while still providing interpretability of distances.

\subsection{MLE Specifications}\label{subsec:models}

The underlying SVC model is \eqref{eq:SVCmodelCHap} where -- as in the simulation study -- we assume exponential covariance functions for all GPs modeling the SVCs. Further, we use regularization for the ranges and variances, i.e. $\mathbb{P}(\rho_j < 1) = 0.05$ and $\mathbb{P}(\sigma_j > 0.3) = 0.05$ in corresponding units. Due to the large number of observations, we apply covariance tapering. The taper range was set to $\rho^\star = 5$ kilometers, as at least half of the observations of the training data have 74 or more neighbors within their taper range, cf.~Figure~\ref{fig:nntaper} in Appendix~\ref{app:taper}. To stay consistent with the definition in the simulation study, we use the same label MLE.r for this model and above described method specifications. 

\subsection{Model Output and Interpretation}\label{sec:mod_inter}

We start with a single MLE.r model output. It was estimated on $\mathcal{S}_{\text{train}, 1}$, cf.~Table~\ref{tab:moving_window}. To provide a reference, we compare it to a classical geostatistical model. That is, the underlying model only contains one GP for the intercept, while all other covariates enter the model only as fixed effects. Though the models differ, we use the same methodology and specification as described for the GP-based SVC model, i.e., profile likelihood optimization using the same regularization and tapering range as previously described. We label it MLE.geo.

The estimated parameters of both the MLE.r and MLE.geo models are given in Table~\ref{tab:MLE_SVCmodel_1}. We note that the estimated mean effects are very similar. Moving to the covariance parameters of the SVC model, note that the estimates are indeed very different for ranges and variances. The bias due to covariance tapering is notable in the range parameters $\widehat{\rho}_j$. The four range parameters exceeding the taper range $\rho^\star $ of 5 kilometers would have an \emph{effective range} of $3\widehat{\rho}_j$ with an exponential covariance function. This would translate to effective ranges of -- in some cases -- well over 200 kilometers, and therefore almost as large as the dimensions of Switzerland. In the context of real estate pricing this would not make sense. But we have to recall that due to tapering all covariance functions are compactly supported on $[0, \rho^\star]$.

\begin{table}[!ht]
	\centering
	\caption{Estimated mean effects, ranges, and variances of the GP-based SVC (MLE.r) and classical geostatistical model (MLE.geo) in the first fold of the moving window validation.}\label{tab:MLE_SVCmodel_1}
	\setcellgapes{2pt}\makegapedcells
	\begin{tabular}{l l | *{3}{r} | *{3}{r} }
	\hline
		&							& \multicolumn{3}{c|}{MLE.r} & \multicolumn{3}{c}{MLE.geo}  \\
	$j$ & SVCs 						& $\widehat{\mu}_j$ & $\widehat{\rho}_j$ & $\widehat{\sigma}_j^2$ & $\widehat{\mu}_j$ & $\widehat{\rho}_j$ & $\widehat{\sigma}_j^2$ \\ \hline 
	1 	& Intercept 				& $8.6448$	& $118.72$  & $0.0210$ & $8.4342$ & $244.70$ & $0.0650$\\
	2	& log \texttt{area} 		& $0.8796$	&  $75.10$  & $0.0010$ & $0.9013$ & -- & -- \\
	3	& \texttt{Z.age} 			& $-0.1707$	&   $4.56$  & $0.0045$ & $-0.1567$ & -- & -- \\
	4	& $\texttt{Z.age}^2$ 		& $0.0272$ 	&	$3.67$  & $0.0002$ & $0.0317$ & -- & -- \\
	5	& \texttt{stand} 			& $0.0825$ 	&  $33.64$  & $0.0009$ & $0.0982$ & -- & -- \\
	6	& \texttt{micro} 			& $0.0820$	&  $39.18$  & $0.0012$ & $0.0988$ & -- & -- \\
	7 	& \texttt{renov} 			& $0.0426$	&   $1.17$  & $0.0034$ & $0.0494$ & -- & -- \\
	8 	& $\texttt{D}_\text{lastQ}$ & $0.0229$ 	&   $0.72$  & $0.0066$ & $0.0249$ & -- & -- \\ \hline
		& Nugget					& --		& 		--& $0.0183$ & --     & -- & $0.0236$ \\ \hline	
\end{tabular}
\end{table}

Further, we notice the relatively large estimates for the variance in the intercept as well as the nugget in the SVC model. Although the covariates are not standardized, they do have a similar range of values that they can take, cf.\ ranges in Table~\ref{tab:covariates}. This leads us to believe that, on the one hand, the mean pricing level is one of the dominant factors when it comes to real estate pricing. On the other hand, the large nugget variance suggests that there exists a high residual variability of apartments within the data set. 

We will now take a look at the visualized SVCs as in Figure~\ref{fig:GWR_facets}, i.e., the intercept, log \texttt{area}, and the ratings of micro locations and standard \texttt{micro} and \texttt{stand}, respectively. These are also the ones with the largest estimated ranges suggesting that the spatial structure will be most prominent in these SVCs. All other covariates SVCs are given in Appendix~\ref{app:SVC_covars}.

\begin{figure}[!ht]
	\centering
	%% 07.../CV6/GWR_facets
	\includegraphics[scale=0.3]{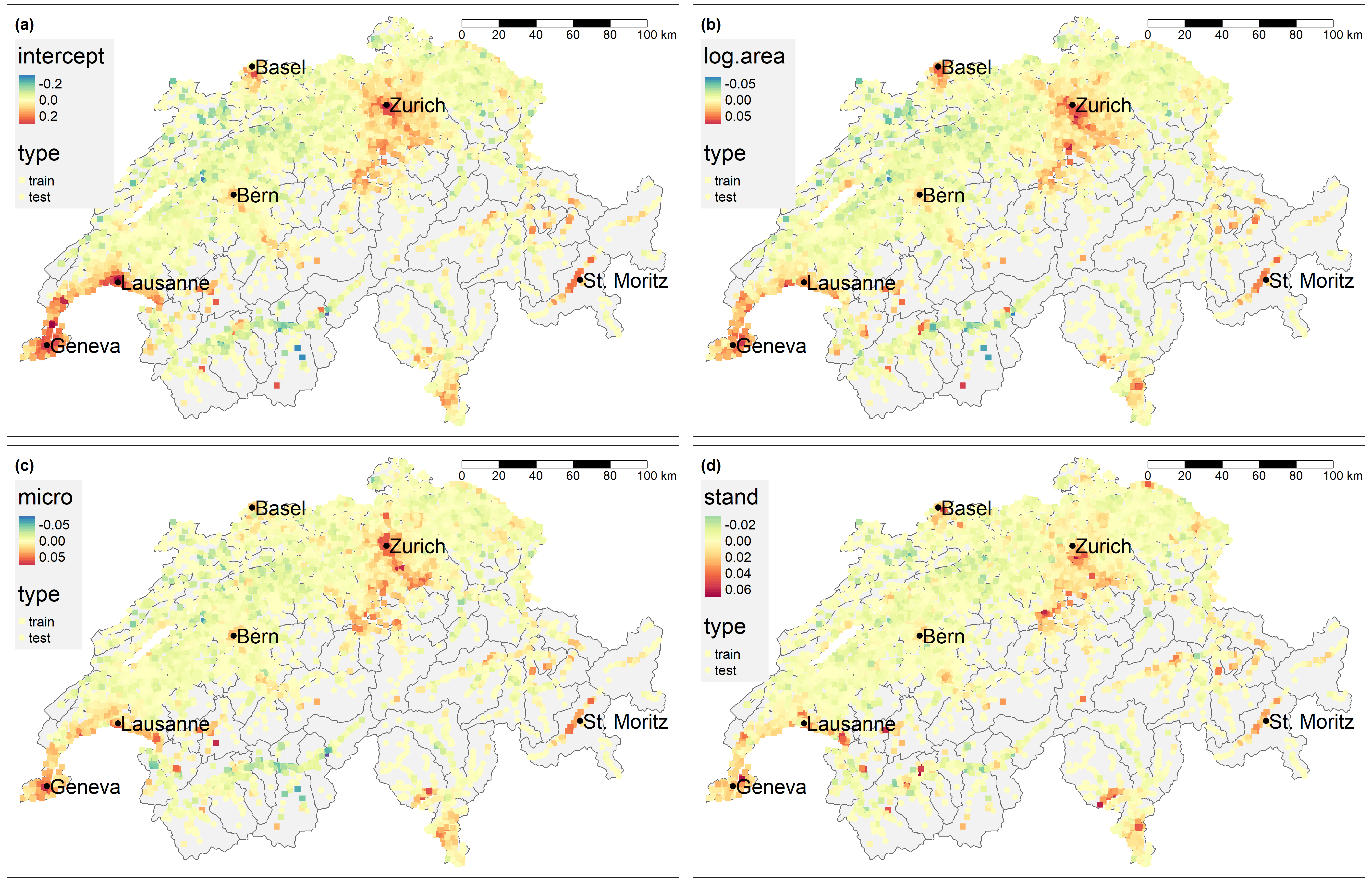}%\includegraphics[scale=0.3]{../figures/use_case/SVC_Q1_facet.png}
	\caption{Selection of estimated SVCs of the model \eqref{eq:SVCmodelCHap} using MLE.r as defined in this section. The four facets show SVCs of (a) the intercept, (b) $\log \texttt{area}$, (c) \texttt{micro}, and (d) \texttt{stand}. The squares indicate training locations $\mathcal{S}_{\text{train}, 1}$, the dots are extrapolations to all other centroids.} \label{fig:MLE_facets}
\end{figure}

Recall that due to their definition as zero-mean GPs, the interpretation of ML-estimated SVCs is different than what we saw in Figure~\ref{fig:GWR_facets}. For GWR the mean effect of each SVC was included. Therefore, one has to interpret Figure~\ref{fig:MLE_facets} as deviations from mean effects, which are given in Table~\ref{tab:MLE_SVCmodel_1}. For the intercept depicted in Figure~\ref{fig:MLE_facets}(a), one can clearly see that the mean apartment prices are highest in the agglomerations of Zurich, Geneva, and Basel as well as the alpine resort Saint Moritz. The mean prices are also higher along the shore line of Lake Geneva starting from the city of Geneva and reaching as far as Lausanne. More rural areas such as between Basel and Bern or in the alpine regions south of Bern have lower mean apartment prices. Qualitatively, the overall picture does not change too much when comparing to Figure~\ref{fig:MLE_facets}(b) to (d), i.e., the SVCs of log \texttt{area}, \texttt{micro}, and \texttt{stand}. There are however some differences. 

For example, along the shore line of Lake Geneva, the effect of \texttt{micro} and \texttt{stand} is smaller compared to the city center of Geneva. However, the mean prizes in Lausanne are relatively high. This suggests that the standard or micro location rating of an apartment in Lausanne is not as important as its location.

Overall, the quality of the ML-estimated SVCs as displayed in Figure~\ref{fig:MLE_facets} seems more plausible and in line with expert knowledge compared to GWR-estimated SVCs in Figure~\ref{fig:GWR_facets}. There are no highly questionable deviations as seen in Figure~\ref{fig:GWR_facets}(a) and (b). Also the estimated ranges vary between the SVCs. \revone{For instance, the SVCs for micro location and standard rating are much more locally pronounced as compared to the intercept and log area.}

\subsection{Predictive Performance}\label{subsec:pred_perf}

\begin{figure}
	\centering
	\includegraphics[width = \textwidth]{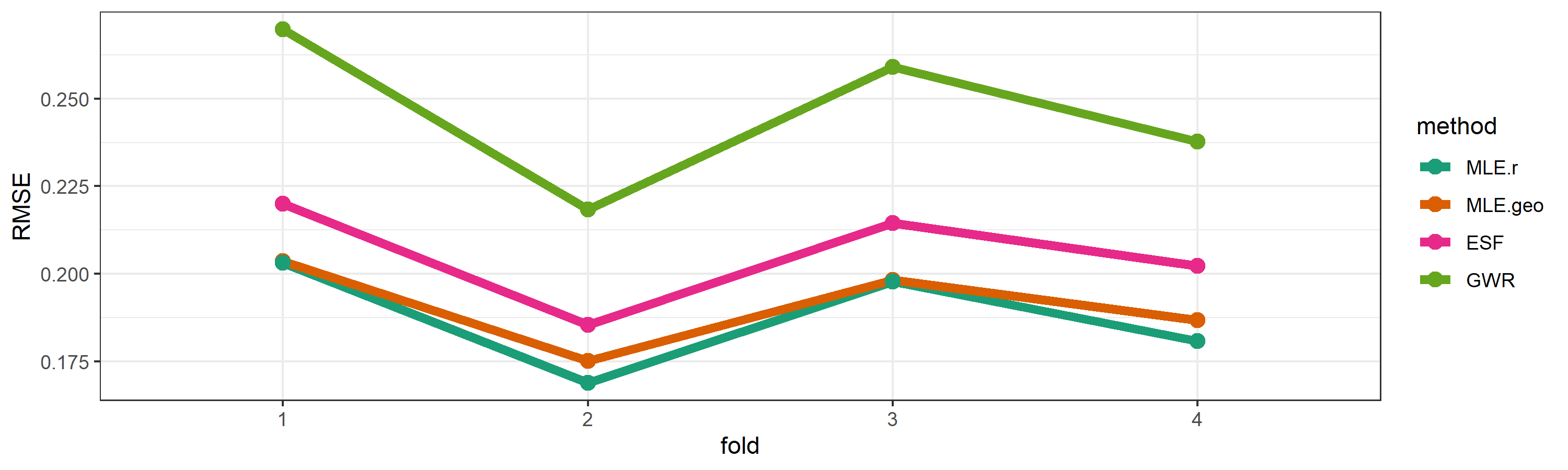}%\includegraphics[width = \textwidth]{../figures/CV6/CV_RMSE_lines.png}
	\caption{$\text{RMSE}^{(m)}_f$ as defined in \eqref{eq:RMSE} of the moving window validation.}\label{fig:CV_RMSE}
\end{figure}

In this last part of the real estate application, we evaluate the predictive performance of MLE.r, MLE.geo, GWR and ESF. The SPDE method cannot be considered as there are too many hyper parameters for \texttt{INLA}. In order to compare the predictive performance of the different approaches, we use the moving window validation setting introduced in Table~\ref{tab:moving_window}. For each fold $f \in \{ 1, 2, 3, 4\}$ and method $m \in \{ \text{MLE.r},\text{MLE.geo},\text{GWR},\text{ESF} \}$, we fit a model $g(\ \cdot \ ; m, f)$ on the data $\mathcal{S}_{\text{train}, f}$ and then make predictions for out-of-sample observations $\iota \in\mathcal{S}_{f}$. \revone{For GWR, this includes an adaptive bandwidth selection as described in Figure~\ref{fig:GWR_facets} for each fold $f$.} Thus, we have $\widehat{y}^{(m, f)}_\iota := g(\textbf{x}_\iota; m, f)$. By doing so, we can compute the prediction errors and the RMSE as follows:
\begin{align}
	e_\iota^{(m)} &:= y_\iota - \widehat{y}^{(m, f)}_\iota \label{eq:error} \\
	\text{RMSE}^{(m)}_f &:= \sqrt{\frac{1}{\left| \mathcal{S}_f \right|}\sum_{\iota \in \mathcal{S}_f} \left(e_\iota^{(m)} \right)^2} \label{eq:RMSE}
\end{align}
We depict the results for the RMSE in Figure~\ref{fig:CV_RMSE}. We find that the GP-based SVC model performs best. Further, it shows that model-based methods outperform ESF and especially GWR throughout all folds. The differences between both MLE for an SVC and a geostatistical model are relatively small.

In the following, we investigate whether the differences are significant by using a mixed effect model describing the RMSE as defined in \eqref{eq:RMSE}. The reference method is the MLE-based SVC model MLE.r defined as the intercept $\alpha_0$. For the methods MLE.geo, ESF, and GWR we include deviations $\alpha_m$ from the reference level $\alpha_0$. Further, we include a random effect for the fold, i.e. a temporal effect, with iid $\zeta_f \sim \mathcal N (0, \sigma_\text{t}^2 )$ as well as an iid noise variable $\varepsilon_{f, m} \sim \mathcal N \left(0, \sigma^2\right)$. The model describing the RMSE is therefore given by
\begin{align}
	\text{RMSE}^{(m)}_f = \alpha_0 + \alpha_m + \zeta_f + \varepsilon_{f, m}.  \label{eq:RMSE_model}
\end{align}

The model has been estimated using the \textsf R package \texttt{lmerTest} \citep{R:lmerTest} and the main results are given in Table~\ref{tab:analysis}, while the whole analysis can be found in the supplementary material. They confirm the conclusions drawn from Figure~\ref{fig:CV_RMSE}. In particular, the RMSE for ESF and GWR is significantly higher than for MLE methods. However, the difference in RMSE between MLE.r and MLE.geo is not significant at 5\% level. 

\subsubsection{Detailed Prediction Error Analysis}

In the following, we perform a detailed prediction error analysis that takes into account that errors of the different methods for the same apartments are correlated. Similarly to model \eqref{eq:RMSE_model}, we analyze the prediction errors as defined in \eqref{eq:error} using a mixed effect model. The independent variables contain a fixed effect for each method which we define with $\gamma_m$. Further, notice that we have repeated measures, since every apartment has been predicted by each method. We can account for this by an iid random effect $\zeta_\iota \sim \mathcal N (0, \sigma_{\textnormal{ap}}^2)$ for the different apartments. Finally, we add iid $\varepsilon_{\iota, m} \sim \mathcal N \left(0, \sigma_m^2\right)$ for the noise with distinct standard deviation for each method. The underlying mixed effect model is 
\begin{align}
	e_{\iota}^{(m)} = \gamma_m + \zeta_\iota + \varepsilon_{\iota, m}.  \label{eq:err_model}
\end{align}

When estimating the model, we expect the estimates of $\gamma_m$ to be near~0, since the predictions should be unbiased. Further, the main focus should be on the estimated standard deviations $\sigma_m$ of the noise, as they indicate the uncertainty of the corresponding methods. The estimation was conducted using the \textsf{R} package \texttt{nlme} \citep{R:nlme} and the main results are given in Table~\ref{tab:analysis}. The estimated fixed effects are in fact close to~0, the estimated standard deviations are increasing throughout the methods indicating a higher precision for predictions under MLE. The standard deviation for the apartment-specific random effect was estimated with $\widehat{\sigma}_{\textnormal{ap}} = \revone{0.1824}$. This again indicates the high variability of apartments that the models have to predict. The full analysis can be found in the supplementary material. 

\begin{table}[!ht]
	\centering
	\caption{Estimated parameters of the RMSE model \eqref{eq:RMSE_model} and error model \eqref{eq:err_model}. The estimated coefficients are given with their corresponding standard errors in parenthesis. Note that in the RMSE model the reference level is the MLE.r method ($\widehat{\alpha}_0 = 0.1877$). Further, the $p$-values for two-sided $t$-tests of $\textnormal{$H_0$: } \alpha_m = 0$ and the estimated standard deviations $\widehat{\sigma}_m$ of model \eqref{eq:err_model} are given.} \label{tab:analysis}
	\begin{tabular}{{l} | *{2}{r} | *{2}{r}}
	\hline
				& \multicolumn{2}{c|}{Model \eqref{eq:RMSE_model}} & \multicolumn{2}{c}{Model \eqref{eq:err_model}} \\
	Methods $m$ & $\widehat{\alpha}_m$ & $p$-value & $\widehat{\gamma}_m$   & $\widehat{\sigma}_m$ \\
	\hline
	MLE.r	& --			 & -- 		&$-$0.0037 (0.0018) & \revone{0.0465} \\
	MLE.geo & 0.0032 \revone{(0.0032)} & \revone{0.3357}	&$-$0.0014 (0.0018) & \revone{0.0516} \\
	ESF		& 0.0179 \revone{(0.0032)} & \revone{0.0003}	&   0.0045 (0.0021) & \revone{0.1202} \\
	GWR		& \revone{0.0586 (0.0032)} & \revone{0.0000}	&\revone{$-$0.0445 (0.0023)}& \revone{0.1674} \\
	\hline
\end{tabular}
\end{table}

\subsubsection{Probabilistic Predictions}

\begin{figure}[!h]
	\centering
	\includegraphics[width = \textwidth]{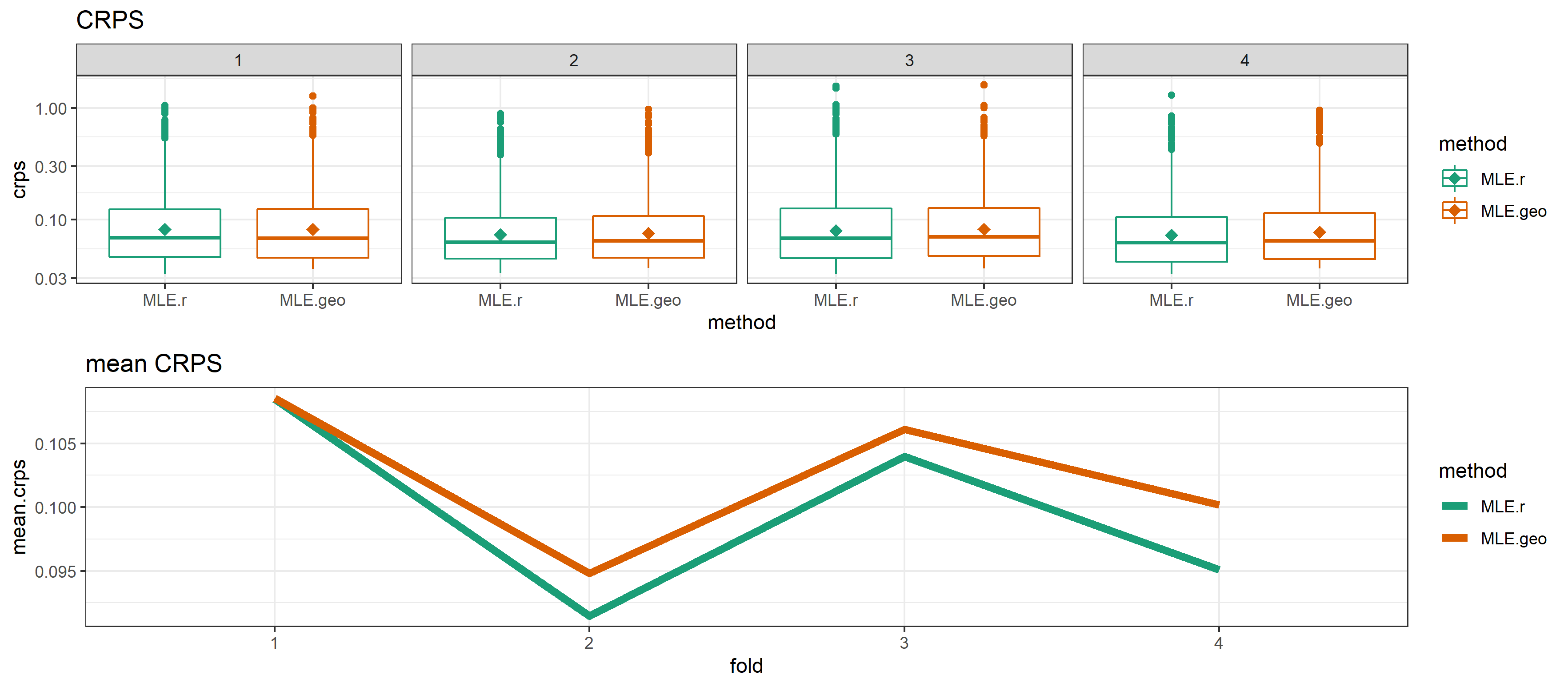}%\includegraphics[width = \textwidth]{../figures/CV6/crps.png}
	\caption{Specific (upper box plots) and mean (lower line plot) CRPS of the MLE methods for the geostatistical model (MLE.geo) and the SVC model (MLE.r) in each fold. Note that the $y$-axis in the upper panel is log-scaled. The values have been computed using the \textsf{R} package \texttt{scoringRules} \citep{R:scroingRules}.}\label{fig:crps}
\end{figure}

The GP-based geostatistical and SVC model allow us to provide a a predictive distribution instead of merely a point prediction. We use the \emph{continuous ranked probability score} (CRPS) proposed by \citet{CRPS2007} to assess the accuracy of the predictive distributions. The results are depicted in Figure~\ref{fig:crps}. Compared to Figure~\ref{fig:CV_RMSE} with the RMSE, one can observe that the SVC model has an advantage with respect to quantifying the uncertainty over the geostatistical model with only having spatially varying intercept. 

\subsubsection{Summary Predictive Performance}

The extensive investigation of predictive performance in the real estate application showed that the MLE method applied on the GP-based models performs considerably better than the other methods, while offering the possibility to quantify the uncertainty of predictions. When it comes to the comparison of the geostatistical to the SVC model, the differences are more subtle. In our application, the gain by modeling and predicting the apartment prices with SVC models instead of geostatistical models is small, whereas the SVC model quantifies uncertainties better. As mentioned in Section~\ref{sec:mod_inter}, this might be due to the location as a dominant factor in real estate mass appraisal.

\section{Conclusion} \label{sec:conclusion}

In this paper, we presented an MLE approach for GP-based SVC models. We empirically validated our approach against other, existing methods for SVC models. This has been done in a simulation study as well as an application with real estate data. To the best of our knowledge, our proposed methodology is the only implemented and currently available method to estimate and make predictions for GP-based SVC models in context of large data where both the sample size and the number of SVCs are large. 

Parameter estimations were shown to be accurate and unbiased when not applying covariance tapering. The predictive performance is in both the simulation study as well as the application among the best. In contrast to not model based approaches such as GWR and ESF it is able to quantify uncertainty by giving predictive variance. 

All GPs were defined by exponential covariance functions and based on Euclidean distances in a two-dimensional domain. However, our proposed method can easily be extended to allow GP-based SVC models defined by individual, non-stationary (anisotropic) covariance functions using other norms on higher dimensional domains $D\subset \mathbb{R}^d$ with $d\geq 3$. Here, MLE could even be augmented to estimate, say, the smoothness. 

We make some final remarks on future work: First and foremost, it would be greatly appreciated to see further comparisons of existing SVC methods on other data sets to see how the predictive performance compares. 

Further, multicollinearity issues with GWR have been raised and one should also investigate if this translates to our methodology. This is in accordance with some SVC selection method that has to be developed to check whether a coefficient is constant or spatially varying.

\section*{Acknowledgment}

We gratefully acknowledge the support by the Swiss Innovation Agency \emph{Innosuisse}, project number 28408.1 PFES-ES. We would like to thank Leonhard Held for his helpful comments on this paper. We also would like to thank Manuel Lehner and Jaron Schlesinger from Fahrl\"ander Partner for their valuable input towards the application with real estate appraisal.

\revone{We appreciate the constructive feedback by the anonymous reviewers and the editor which improved the quality of this work.}

\section*{Supplementary Material}

The code for computation is available online under \url{https://git.math.uzh.ch/jdambo/open-access-svc-paper}. Additionally, we provide the following supplementary material. 

\begin{description}

\item[Appendix:] Additional results and figures. (PDF)

\item[\textsf{R} package \texttt{varycoef}:] \textsf{R} package by \citet{R:varycoef} containing the routines of the MLE method for SVC models described in this article. (GNU zipped tar file) 

\end{description}

\bibliographystyle{../../my_bib/chicago}

\bibliography{../../my_bib/mybib}

%\bibliography{mybib}

\newpage

\appendix

\begin{center}
{\large \bf Appendix to: \\ \protect Maximum Likelihood Estimation of Spatially Varying Coefficient Models for Large Data with an Application to Real Estate Price Prediction}
\end{center}

\section{Further Results: Simulation Studies}

In this section, we show all the results of the simulation study that were only mentioned or not shown at all in the main article.

\subsection{Simulation 2: Number of Observations is \texorpdfstring{$n' = 10{,}000$}{n10000}} \label{appendix:sim2}

\revone{As Figure~\ref{fig:sim1_RMSE} for Simulation~1, we give the RMSE results for Simulation~2 in the following.}

\begin{figure}[!h]
	\centering
	\includegraphics[scale=0.6]{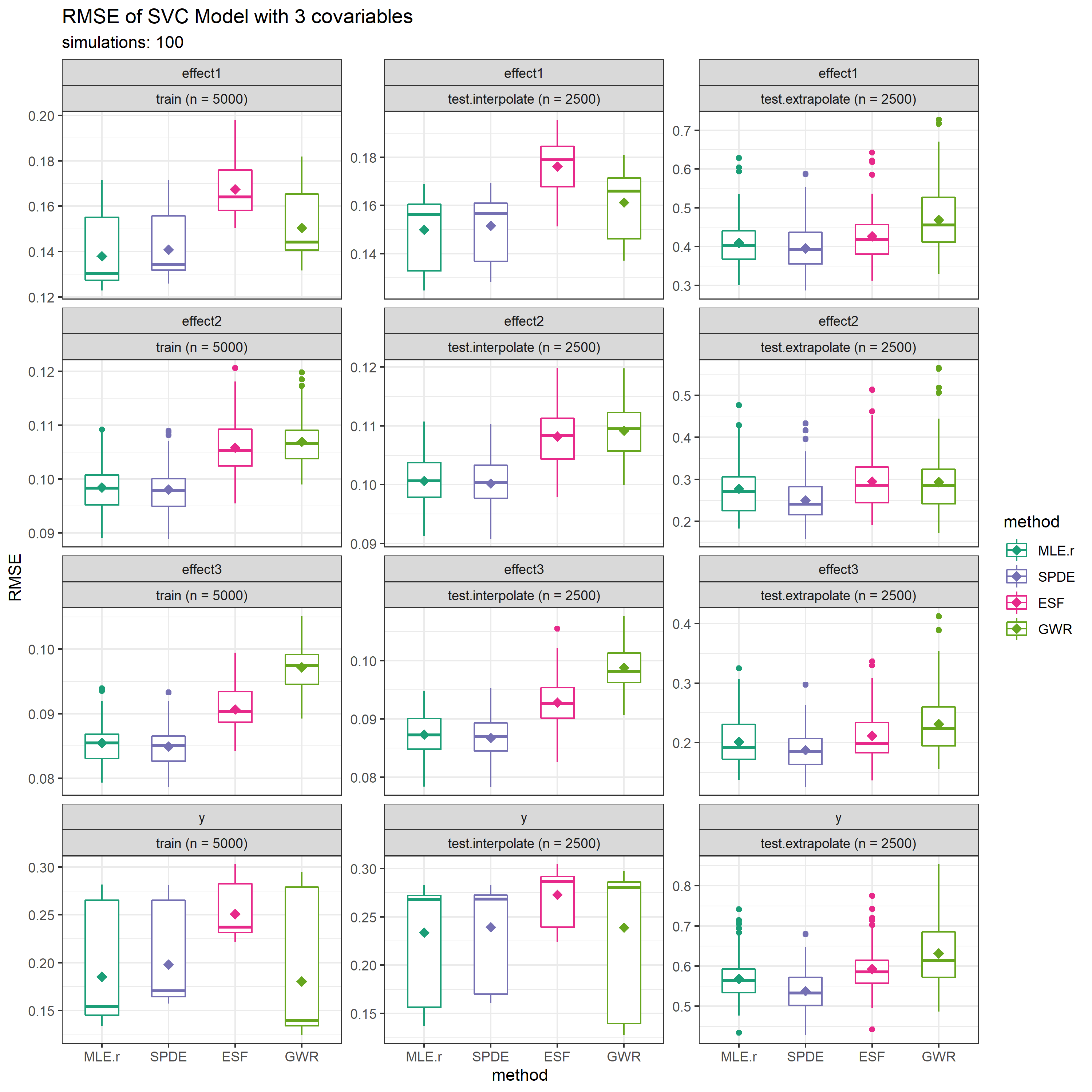}%../code/simulation/plots/sim2_RMSE_all_revision.png
	\caption{\revone{Box plots of RMSE between true $\bm \beta_j$ and $\widehat{\bm \beta}_j$ as well as between $\mathbf y$ and $\mathbf{\widehat{y}}$ for Simulation~2 in corresponding partition, see \eqref{eq:RMSEbeta} and \eqref{eq:RMSEy}. The diamonds indicate the means. Note the different ranges of the $y$-axis.}} \label{fig:sim2_RMSE}
\end{figure}

In Figure~\ref{fig:sim2_RMSE}, \revone{one can see that the model-based approaches are best in the first two columns, i.e., for training and spatially interpolating predictions. For spatial extrapolation, the RMSEs are quite similar with the ranking SPDE, MLE.r, ESF, and GWR (best to worst). For the response's RMSE \eqref{eq:RMSEy}, we observe that GWR's errors spread a wide range in the first two columns.}

\begin{figure}[!h]
	\centering
	\includegraphics[scale=0.6]{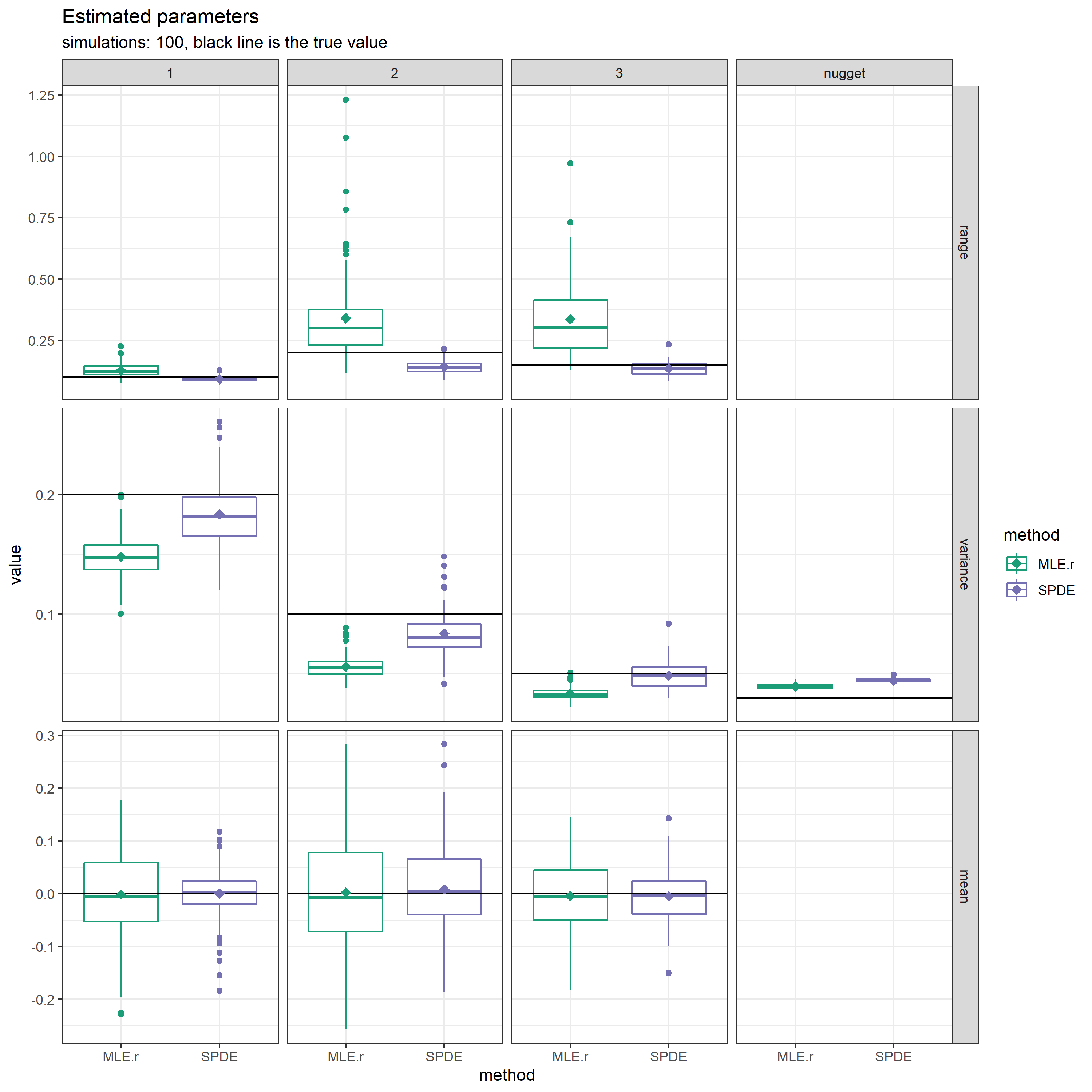}%../code/simulation/plots/sim2_parameters_revision.png
	\caption{Estimated covariance and mean parameters of all methods of the 100 simulations given as box plots in Simulation~2. The diamonds indicate the means. True values are given by black lines.}\label{fig:sim2_parameters_all}
\end{figure}

\FloatBarrier

\subsection{Simulation 3: Number of SVC is \texorpdfstring{$p = 10$}{p10}} \label{appendix:sim3}

In this simulation, \revone{MLE.r} is the only method to estimate the parameters of the SVC model. The accuracy in doing so is surprisingly good. The mean parameters were all estimated without bias and with a deviation proportional to the corresponding variance of the respective GP. The ranges of most SVC are slightly biased in the sense that they are underestimated. This probably leads to underestimated variances, too. 

\begin{figure}[!h]
	\centering
	\includegraphics[scale=0.6]{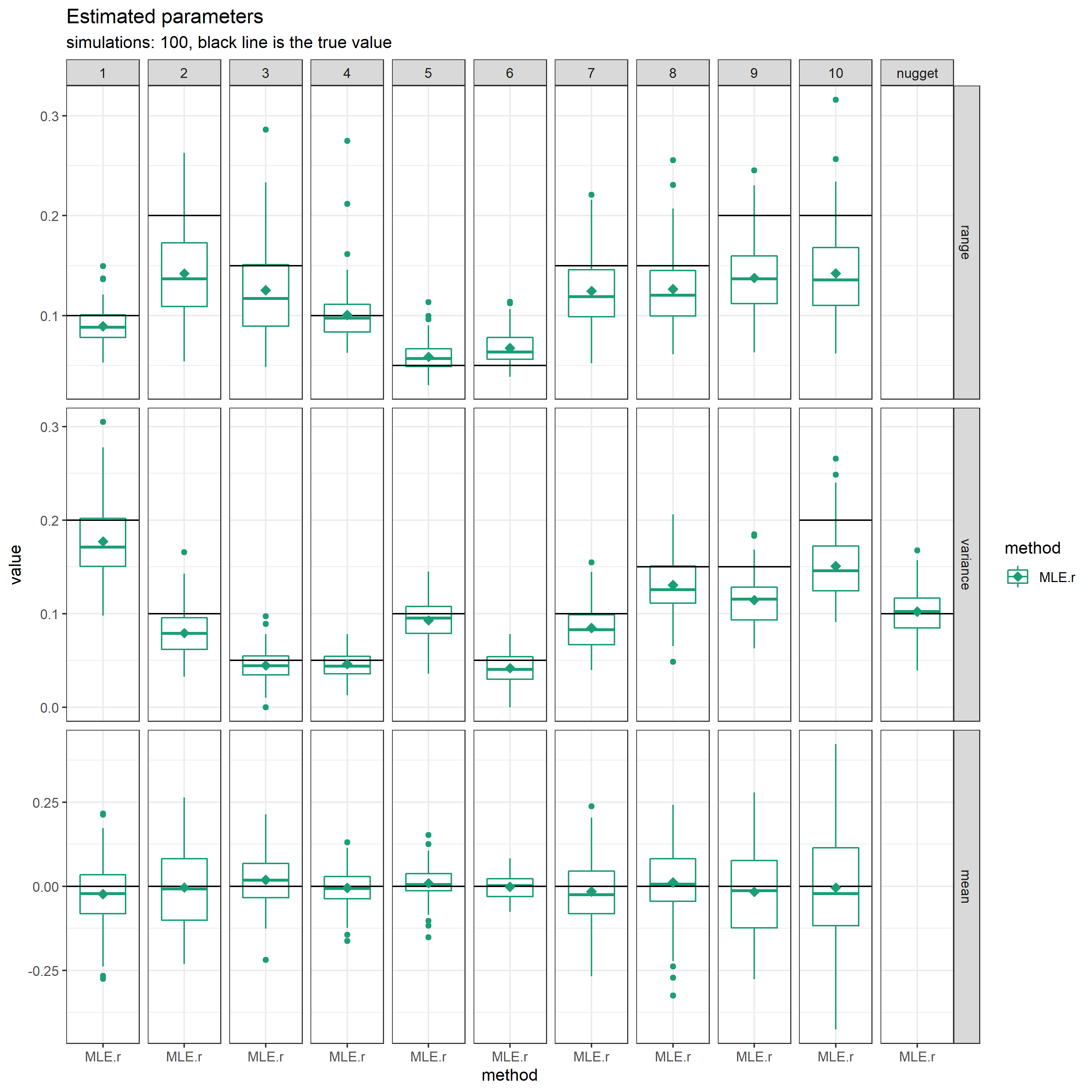}%../code/simulation/plots/sim3_parameters_revision.png
	\caption{\revone{Estimated covariance and mean parameters of all methods of the 100 simulations given as box plots in Simulation~3. The diamonds indicate the means. True values are given by black lines.}} \label{fig:sim3_parameters}
\end{figure}

\FloatBarrier

Over the next 3 pages we show the RMSE for Simulation~3.

\begin{figure}[!h]
	\centering
	\includegraphics[scale=0.5]{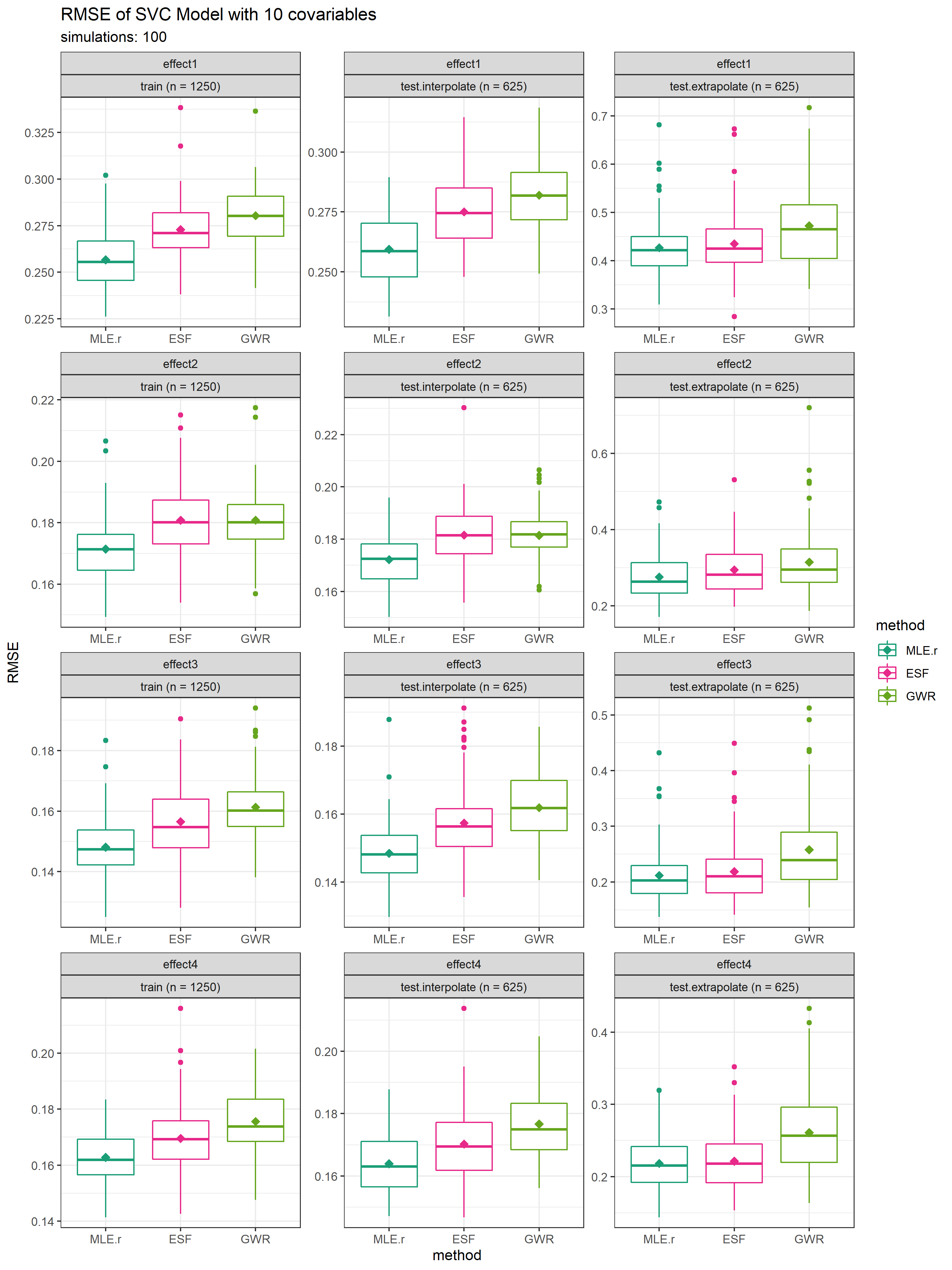}%../code/simulation/plots/sim3_RMSE_all_1_revision.png
\end{figure}

\begin{figure}[!h]
	\centering
	\includegraphics[scale=0.5]{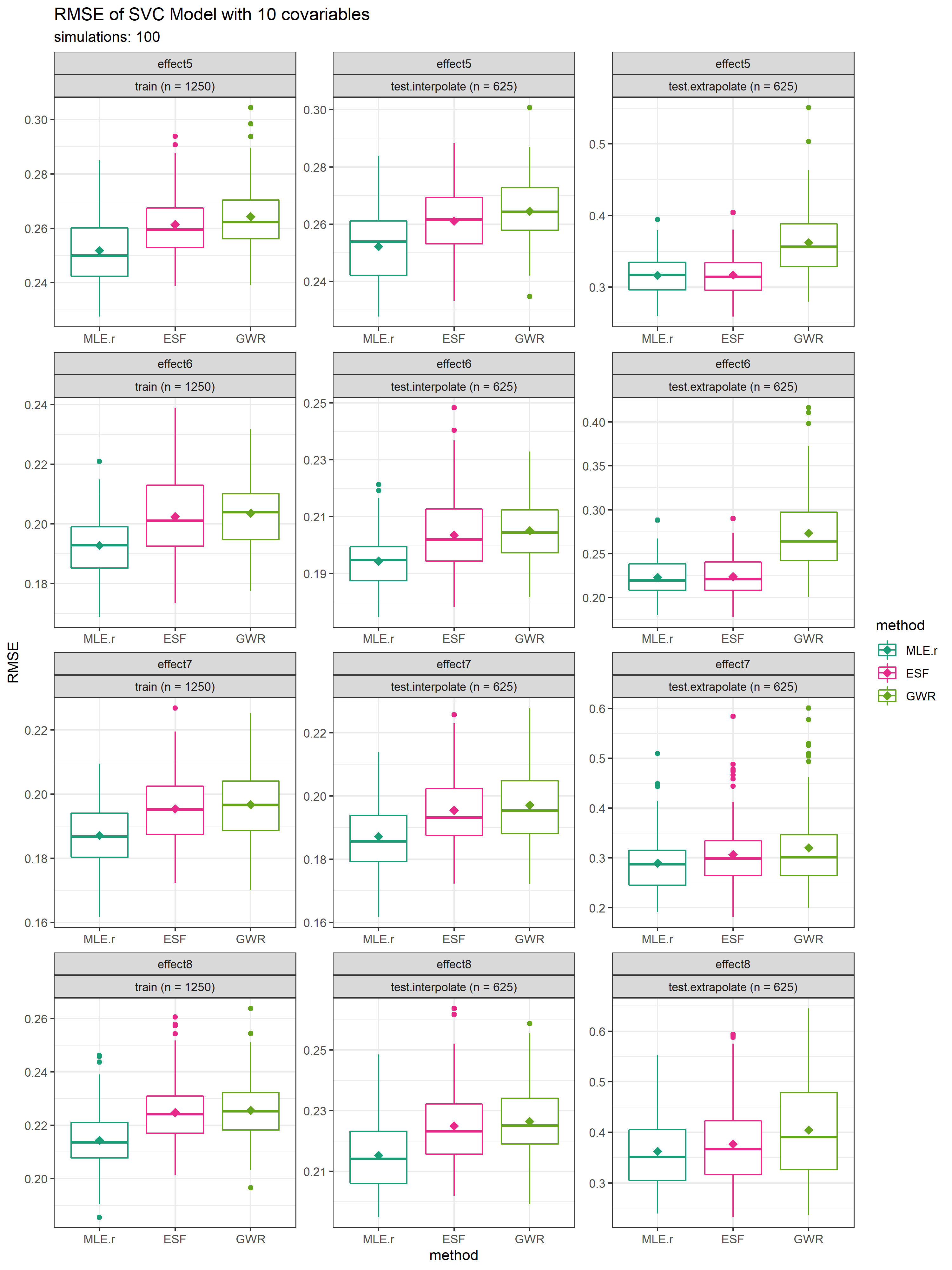}%../code/simulation/plots/sim3_RMSE_all_2_revision.png
\end{figure}

\begin{figure}[!h]
	\centering
	\includegraphics[scale=0.5]{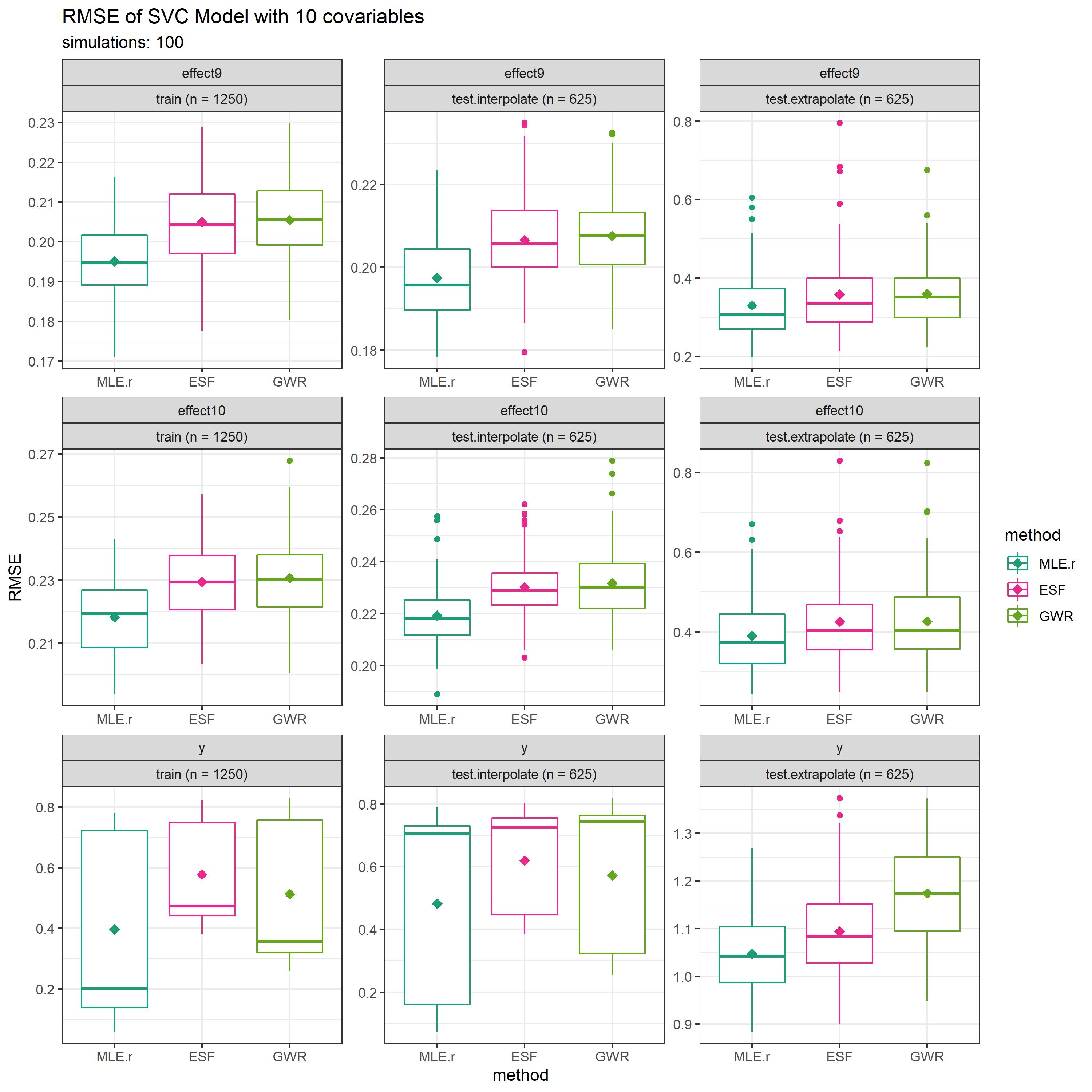}%../code/simulation/plots/sim3_RMSE_all_3_revision.png
	\caption{Box plots of RMSE between true $\bm \beta_j$ and $\widehat{\bm \beta}_j$ as well as between $\mathbf y$ and $\mathbf{\widehat{y}}$ for Simulation~3 in corresponding partition, see \eqref{eq:RMSEbeta} and \eqref{eq:RMSEy}. The diamonds indicate the means. Note the different ranges of the $y$-axis.} \label{fig:sim3_RMSE_all}
\end{figure}

\FloatBarrier

\section{Choice of Taper Range} \label{app:taper}

\begin{figure}[!h]
	\centering
	\includegraphics[width=\textwidth]{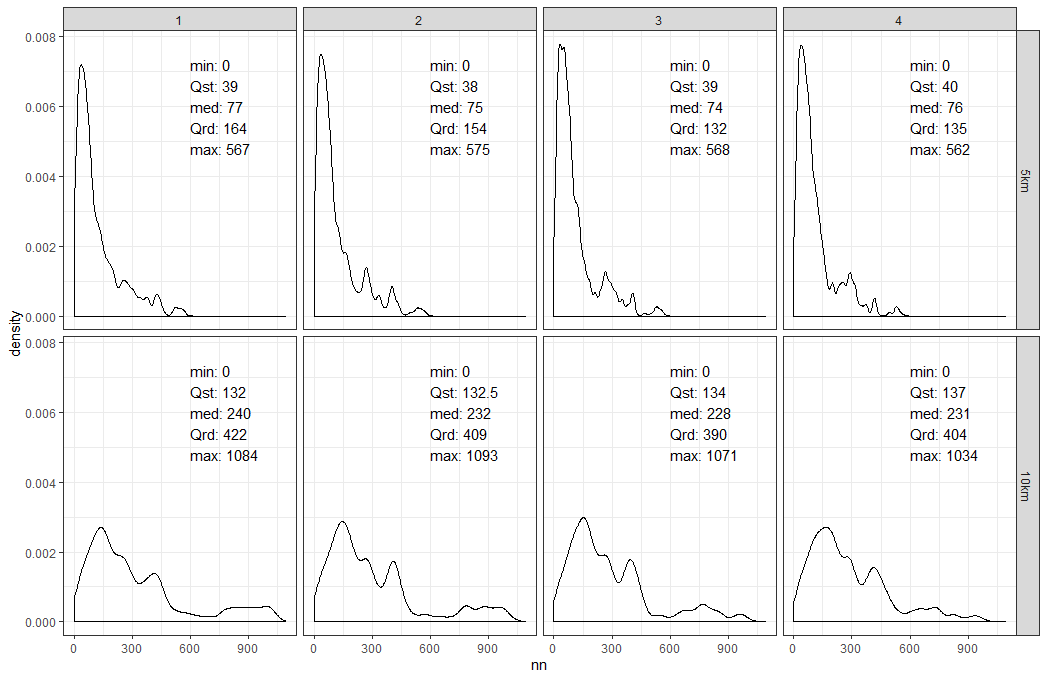}%../figures/CV6/nn_taper.png
	\caption{The densities of $nn_i^{\rho^\star}$ for $\rho^\star \in \{5, 10\}$ and each fold are given in corresponding facets. Additionally, we give the respective quartiles.} \label{fig:nntaper}
\end{figure}

In order to make a good choice on the taper range for MLE.geo and MLE.r, we have to investigate how many neighbors each observation with a given taper range has. Thus, we compute the \emph{number of neighbors} for an observation $i$ as
\begin{align*}
	nn_i^{\rho^\star} := \sum_{j\neq i} 1_{\left\{ \| s_i - s_j \| \leq \rho^\star\right\} },
\end{align*}
where $1_{A}$ is the indicator function which is 1 if $A$ holds true and 0 otherwise. We computed $nn_i^{\rho^\star}$ for two taper ranges $\rho^\star \in \{5, 10\}$. The density and some summary statistics are depicted in Figure~\ref{fig:nntaper} for each fold. The median of $nn_i^{\rho^\star = 5}$ is at least 74 over all folds and the first quartile is at least 38 over all folds, respectively. The maximum does not exceed 575, meaning that there is no observation with more than 575 neighbors within a radius of 5 kilometers. In the case of $\rho^\star = 10$ we see that the maximum of $nn_i^{\rho^\star = 10}$ exceeds 1000 in all folds, which is why we choose a 5 kilometer range over a 10 kilometer range. 

\section{Estimated SVC} \label{app:SVC_covars}

Finally, we show the other estimated SVC for fold $f = 1$, as we did in Figure~\ref{fig:MLE_facets} for the MLE.r method. These are the deviations $\widehat{\bm \eta}_j$ from the corresponding mean $\mu_j$, which added result in a single SVC for a covariate. 

\begin{figure}[!h]
	%% 07.../CV6/GWR_facets
	\centering
	\includegraphics[scale=0.3]{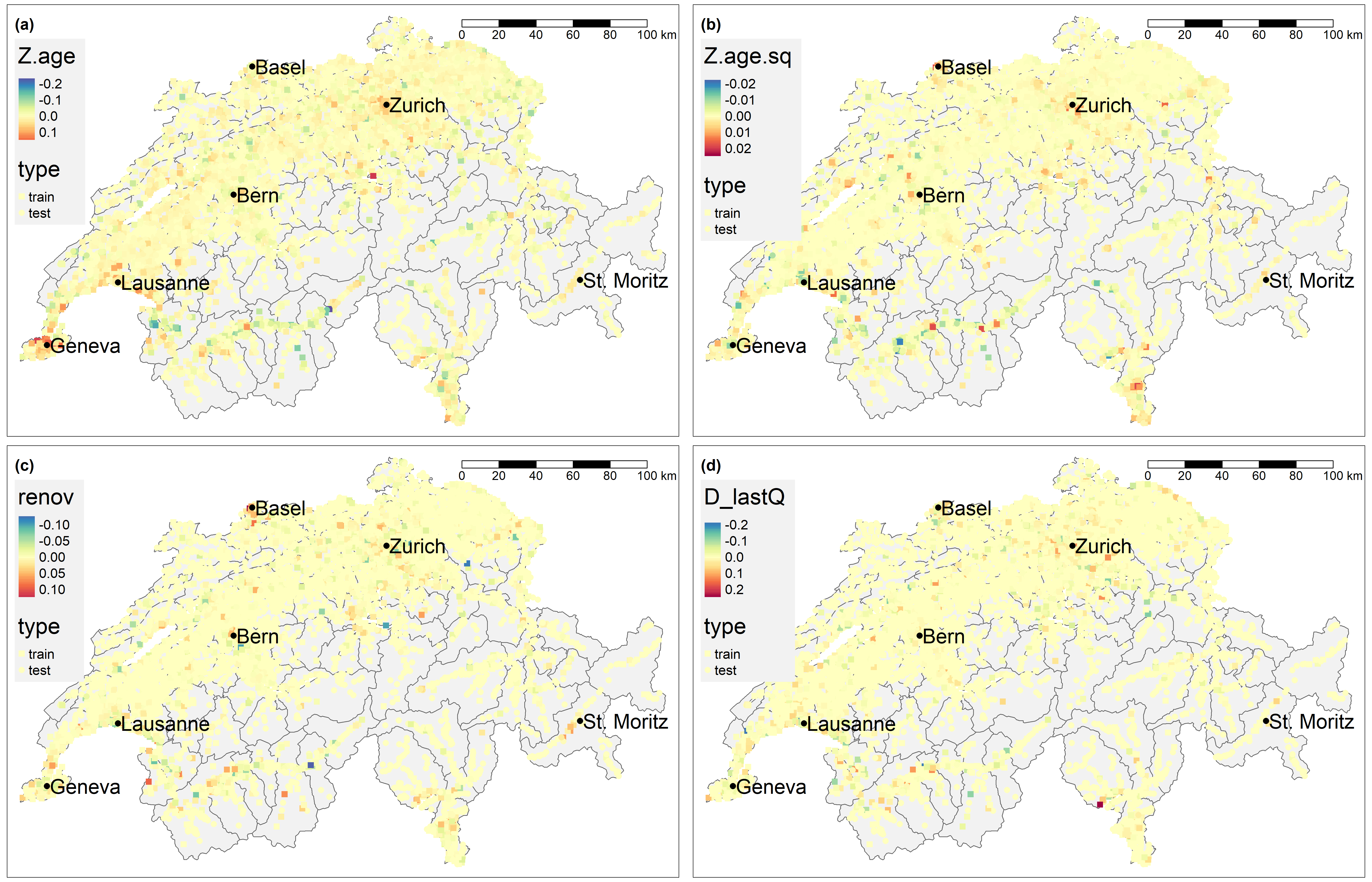}%../figures/use_case/SVC_Q1_facet2.png
	\caption{ML-estimated, further SVCs of the model \eqref{eq:SVCmodelCHap} using MLE.r as defined in this Section~\ref{sec:CHap}. These estimated SVCs are supplementary to the ones depicted in Figure~\ref{fig:MLE_facets}. This figure shows (a) $\widehat{\bm \eta}_3$ for covariate \texttt{Z.age}, (b) $\widehat{\bm \eta}_4$ for covariate $\texttt{Z.age}^2$, (c) $\widehat{\bm \eta}_7$ for covariate \texttt{renov}, and (d) $\widehat{\bm \eta}_8$ for covariate $\texttt{D}_\textnormal{lastQ}$. The squares indicate training locations $\mathcal{S}_{\text{train}, 1}$, the dots are extrapolations to all other centroids.} \label{fig:SVC2_facets}
\end{figure}

\paragraph{Age:} Of the four SVC yet not discussed in this paper, both the linear as well as quadratic effect of the age covariate have the most pronounced spatial structure, cf.\ Figure~\ref{fig:SVC2_facets}(a) and (b). One can see that the effect of age in agglomerations is very different from more rural areas. This is consistent with the assumption that -- marginally -- in rural areas one expects pure depreciation of an apartment with increasing age, while within city centers newly built and old (apartments with year of construction prior to 1920) are higher priced. 

\paragraph{Renovation Rating and Last Quarter:} Both of theses covariates do not have very strong spatial structures, cf.\ Figure~\ref{fig:SVC2_facets}(c) and (d).

\end{document}